\documentclass[12pt,preprint]{aastex}
\def\Msun{M$_{\odot}$}

\newcommand {\tighttable}{\def\baselinestretch{1.0}}

\newcommand {\ergscmA}{${\rm erg\,s}^{-1}\,{\rm cm}^{-2}\,{\rm \AA}^{-1}$}

\newcommand {\ergs}{${\rm erg\,s}^{-1}$}
\newcommand {\cc}{$\rm cm^{-3}$}
\newcommand {\cl}{$\rm cm^{-2}$}
\newcommand {\kms}{${\rm km\,s}^{-1}$}
\newcommand {\ltsim}{\raisebox{-.5 ex}{$\;\stackrel{<}{\sim}\;$}}

\newcommand {\HST}{{\it HST}}
\newcommand {\lya}{Ly~$\alpha$}

\newcommand {\h}{\ion{H}{1}}

\def\arcsecpoint{\ifmmode ''\!. \else $''\!.$\fi}
\newcommand {\he}{\ion{He}{2}}
\newcommand {\heiii}{\ion{He}{3}}
\newcommand {\oi}{\ion{O}{1} $\lambda 1304$}

\slugcomment{To appear in {\it The Astrophysical Journal}}

\newcommand {\obj}{SDSS1253+6817}
\shorttitle{Intergalactic helium}
\shortauthors{Zheng et al.} 
\received{2014 April 18}
\accepted{2015 April}
\articleid{MS 94602}{}
\begin{document}
\title{Characteristics of He~{\sc ii} Proximity Profiles}  
\author{
Wei Zheng\altaffilmark{1},
David Syphers\altaffilmark{2},
Avery Meiksin\altaffilmark{3},
Gerard A. Kriss\altaffilmark{4},
Donald P. Schneider\altaffilmark{5,6},
Donald G. York\altaffilmark{7},
and
Scott F. Anderson\altaffilmark{8}
} 
\altaffiltext{1}{Department of Physics and Astronomy, Johns Hopkins 
University, Baltimore, 3701 San Martin Dr., MD 21218, USA} 
\altaffiltext{2}{Physics Department, East Washington University, Science 154, Cheney, WA 99004, USA} 
\altaffiltext{3}{Scottish Universities Physics Alliance (SUPA); Institute for Astronomy, 
 University of Edinburgh, Royal Observatory, Edinburgh EH9 3HJ, United Kingdom}
\altaffiltext{4}{Space Telescope Science Institute, Baltimore, 3700 San 
  Martin Dr., MD 21218, USA}
\altaffiltext{5}{Department of Astronomy and Astrophysics, 525 Davey Lab., The Pennsylvania 
State University, University Park, PA 16802, USA} 
\altaffiltext{6}{Institute for Gravitation and the Cosmos, The Pennsylvania 
State University, University Park, PA 16802, USA} 
\altaffiltext{7}{Department of Astronomy and Astrophysics and the Fermi 
Institute, 5640 S. Ellis Avenue, The University of Chicago, Chicago, IL 60637, USA}
\altaffiltext{8}{Department of Astronomy, University of Washington, 3910 15th Ave NE,
Seattle, WA 98195, USA}
\begin{abstract}

The proximity profile in the spectra of $z\approx 3$ quasars, where fluxes 
extend blueward of the He~{\sc ii} Ly~$\alpha$ wavelength $304 (1+z)$ \AA, 
is one of the most important spectral features in the study of the intergalactic 
medium. Based on the \HST\ spectra of 24 \he\ quasars, we find that the majority of them display a proximity profile, corresponding to an ionization radius as large as 20 Mpc in the source's rest frame. In comparison with those in the \h\ spectra 
of the quasars at $z \approx 6$, the \he\ proximity effect is more 
prominent and is observed over a considerably longer period of reionization.
The \he\ proximity zone sizes decrease at higher redshifts, particularly at $z>3.3$. This trend is similar 
to that for \h, signaling an onset of \he\ reionization at $z\gtrsim 4$. 

For quasar SDSS1253+6817 ($z=3.48$), the He~{\sc ii}  absorption trough displays a 
gradual decline and serves as a good case for modeling the \he\ reionization.
To model such a broad profile
requires a quasar radiation field whose energy distribution between 4 and 1 Rydberg  is considerably 
harder than normally assumed. The UV continuum of this quasar is indeed exceptionally steep, and
the \he\ ionization level in the quasar vicinity is higher than the average level in the 
intergalactic medium. These results are evidence that a very hard EUV continuum
from this quasar produces a large ionized zone around it.
 
Distinct exceptions are the two brightest \he\ quasars at $z\approx 2.8$, for which no significant 
proximity profile is present, probably implying that they are very young.

\end{abstract}

\section{INTRODUCTION}\label{sec_intr}
The epoch at redshift $\approx 3$ marks the peak of quasar formation 
\citep{richards} as well as the reionization of the intergalactic helium 
\citep[][M09 hereafter]{meiksin09}. 
As the reionization of singly ionized helium requires high-energy photons above 4 Rydberg, the hard UV 
background radiation field in the vast intergalactic space, commonly referred to as ``metagalactic'', at 
$z\approx 3$ is believed to be photons originating from quasars instead of hot stars \citep{haardt,meiksin05}.
Because helium is difficult to ionize and readily recombines, its opacity at this redshift range is 
considerably higher than that of hydrogen.

Most of our knowledge of the intergalactic helium at these high redshifts 
is based upon a few bright quasars with clear lines of sight: QSO0302$-$003 \citep{jakobsen,hogan,heap}, 
HS1700+6416 \citep{dkz,fechner} and HE2347$-$4342 \citep{gak,smette,shull04,zheng04,shull}.
Thanks to the Sloan Digital Sky Survey \citep[SDSS,][]{york} and Galaxy Evolution Explorer
\citep{martin}, the number of known quasars with unobscured sightlines to their \he\ \lya\ 
wavelengths has increased dramatically from only three in mid-1990s to more than 50 \citep{syphers0,syphers1, 
worseck}. Of these quasars the highest redshift is 3.93 \citep{syphers1}. 
The Cosmic Origins Spectrograph \citep[COS,][]{green} instrument aboard the \HST\ has greatly enhanced 
our ability to probe multiple lines of sight and investigate the \he\ absorption features at
higher signal-to-noise (S/N) ratios. In the five years since 
the COS installation, more than two dozen quasars have been confirmed with the \he\ \lya\ features.

The ensemble of spectra of these quasars is now sufficiently large that we can begin to overcome 
the cosmic variance and investigate the 
reionization process over a significant range of redshift. It is now well established that the
Gunn-Peterson effect \citep{gp} in quasar spectra becomes significant only at 
$z\gtrsim 2.7$ for \he\
\citep{dkz,reimers,anderson} and at $z \gtrsim 5.5$ for \h\ 
\citep{becker,fan}, signaling the end of the respective reionization epochs.
The intergalactic-medium (IGM) reionization process is lengthy, and little is 
known about the IGM evolution at much higher redshifts, {\it i.e.}, $z \gtrsim 3.5$ 
for helium, $z\gtrsim 7$ for hydrogen.

The ionization properties of the IGM change significantly near luminous quasars, whose radiation 
enhances the ionization level in  their vicinity. This ``proximity effect'' was first found in quasars at $z\approx 3$
where the number of \h\ \lya\ forest lines in their optical spectra declines in the quasar 
vicinity \citep{murdoch,carswell,tytler2}. 
At a high IGM opacity, the proximity effect results in an expected  residual flux blueward of the
\he\ \lya\ wavelength \citep{zheng95,giroux,madau00}. Such an absorption 
profile has been observed in the \h\ spectra of most quasars at 
$z>5.7$ \citep{white,carilli}. For helium, the presence of proximity 
profiles was ambiguous, as there were only a handful of known \he\ quasars, and some of 
them do not display a proximity profile.

The \he\ \lya\ absorption troughs may be complex \citep{madau00}, as their shapes are affected by 
several factors: a proximity profile blueward of the \he\ \lya\ wavelength, 
a damped absorption profile that extends to the wavelengths redward of it, and possibly another 
damped absorption  profile from an associated absorber. In this paper, we report our analysis of 
the proximity effect using a large sample of quasars with confirmed \he\ \lya\ features.
The distances quoted in this paper are proper distances in the quasar's rest frame, calculated 
using the codes of \cite{hogg}, with the standard cosmological 
parameter values:\ $\Omega_M=0.3$, $\Omega_\Lambda=0.7$ and $h=H_0/100\,{\rm
km\,s^{-1}\,Mpc^{-1}}=0.70$. 
\section{OBSERVATIONS AND ANALYSES}\label{sec_data}

We investigate the \HST\ spectra of 24 quasars in which the \he\ \lya\ 
absorption feature is present at the anticipated wavelength and is not severely contaminated 
by geocoronal emission. Our database
consists of three parts: (1) COS spectra of four quasars at $z>3.4$ from our program 
(GO 12249: PI Zheng), see Table \ref{tbl-1};
(2) archival COS spectra of seven quasars (GO 11742, 13013: PI Worseck); and
(3) published spectra of 13 quasars, taken with COS and STIS.
Six other quasars are excluded because their \he\ break is contaminated by 
geocoronal \lya\ or \ion{O}{1} $\lambda 1302$ emission. 
Table \ref{tbl-2} lists the eleven sources in parts 1 and 2; and Table \ref{tbl-3} 
lists the 13 sources in part 3. 

Optical spectra of these quasars are useful as they provide information of the \h\
absorption counterparts. We retrieved 15 SDSS spectra of the \he\ quasars,  the
VLT/UVES spectra of quasars PKS1935$-$692 and HE2347-4342, and the Keck/HIRES spectra of QSO0302$-$003 
and HS1700+6416. 
In total, we have the optical spectra of 17 quasars, among which four are at high resolution.

\subsection{COS Spectra of Four Quasars}\label{sub_cos}

We carried out \HST/COS spectroscopy of four quasars: SDSS1253+6817, SDSS1319+5202, 
SDSS1711+6052, SDSS2346$-$0016 (GO program 12249). 
They were selected as a bright sample (the UV continuum level at $\gtrsim 
5\times 10^{-17}$ \ergscmA) at $z>3.4$ with confirmed \he\ \lya\ absorption breaks 
with the ACS prism data \citep{zheng08,syphers0,syphers1}. The new 
COS data provide considerably higher S/N ratios and spectral resolution.
The \HST/COS observations of these four quasars were performed between 2010 November 
and 2011 December. The G140L spectra cover a 
wavelength range of approximately 1100-2150 \AA\ with a pixel scale of 0.08 \AA\ and 
resolution $R\approx 2000$. 
The source fluxes were extracted using a small slit width of 25 pixels, which is considerably smaller 
than the default size of 57 pixels \citep[0\arcsecpoint 59, ][]{syphers4, syphers5}. 
To determine the level of geocoronal-line contamination and scattered light, we reduced portions 
of the orbital night separately and compared with the full set. 
Only in one case (SDSS2346$-$0016) the airglow is significant even during the orbital 
night \citep{feldman}, and the dataset taken in 2011 December has to be excluded. 

The spectrum of SDSS1319+5202 reveals a Lyman-limit system (LLS) at  $z\approx 0.7$ ($\sim 1550$~\AA\ in the observed frame), 
and the source flux below 1550~\AA\ is considerably suppressed. 
Based on low-resolution prism data ($R\simeq 105$ at 1460 \AA), 
we suggested an associated damped absorption system 
whose effect extends redward of the \he\ \lya\ wavelength \citep{zheng08}. With the improved wavelength accuracy
of our higher-resolution COS spectrum, the 
absorption profile is confirmed to straddle the \he\ \lya\ wavelength and cannot be explained by a simple
proximity profile. 
The COS spectra of these four quasars are shown in Figure \ref{fig-cos}. 

To study the absorption profiles, it is essential to accurately determine the quasar systemic redshifts.
We retrieved the optical spectra of these quasars from the SDSS and used the 
IRAF task {\tt specfit} \citep{specfit} to fit both the optical and UV spectra with 
multiple components. For the COS spectra, a power-law continuum, and a \he\ \lya\ 
emission line were used to fit the data in wavelength redward of the \he\ \lya\ break.
For the optical spectra, we used a power-law continuum, multiple emission lines, including 
\lya\ (narrow + broad), \ion{N}{5}, \ion{O}{1}, \ion{Si}{4}, \ion{C}{4}, \ion{He}{2} \citep{vb}, and 
a set of absorption templates that simulate the accumulated IGM absorption and the proximity profiles.
The systemic redshifts were verified using the \ion{O}{1} $\lambda 1304.22$ emission line as low-ionization lines are believed to be less affected by the local motion of broad-line 
regions. For our sample, the redshifts are generally consistent with those in the SDSS 
Data Release Seven Quasars Catalog \citep[DR7,][]{dr7}, except for SDSS2346$-$0016. 
We obtained a spectrum of SDSS2346$-$0016 using the TripleSpec instrument 
\citep{wilson} on the ARC (the Astrophysical Research Consortium) 3.5-m telescope. 
The observations were made on 2011 November 10 and 16, under an ambient temperature 
of 1-3 C, and the total integration time was 10 hours.
As shown in Figure \ref{fig-nir}, the [\ion{O}{3}]~$\lambda$5007 emission is 
not visible. The \ion{Mg}{2} and H$\beta$ emission lines, while 
unusually weak, yield a redshift consistent with the \ion{O}{1} line. 
The COS spectra are of sufficient spectral resolution to confirm that
that our measurements of the proximity profile are not 
affected noticeably by the small difference in quasar redshifts.  

\subsection{Archival COS Spectra of Seven Quasars}\label{sub_arc}

We retrieved the archival COS/G140L spectra of 15 other quasars (GO 11742 and 13013)
via the Mikulski Archive for Space Telescopes 
and processed with the standard pipeline {\tt calcos} \citep[][ v2.19]{hodge}. 
The difference between the pipeline extraction and our narrow-window extraction (see \S \ref{sub_cos}) is mainly in the 
residual flux level at wavelengths far away from the quasar, but this difference has no effect to our study.
For QSO0233$-$0149, the geocoronal  \ion{O}{1} $\lambda 1302$ line is close to the \he\ \lya\ 
wavelength, therefore we used the orbital-night portion of the data.
We also used the night portion of the QSO0916+2405 data as 
geocoronal \ion{O}{1} $\lambda 1356$ emission is near its \he\ proximity profile. 
Four of these quasars were excluded because of severe contamination by geocoronal emission.
Four more were excluded as the flux breaks in their spectra are observed at 
wavelengths considerably longer than that for \he\ \lya. Most likely,
these are attributed to partial \he\ LLSs along the lines of sight.
The details of the seven archival spectra used are listed in the lower parts of Tables \ref{tbl-1} and \ref{tbl-2},
below the details for our prime COS spectra listed at the top of the tables.  
Figure~\ref{fig-cos2} plots the spectra of seven quasars that we include in our sample, with contaminating lines 
flagged with the Earth symbols.

\subsection{Archival Data in the Literature}\label{sub_liter}

A number of the COS spectra of \he\ quasars have been published \citep{shull, syphers4, syphers5}, most of which were observed with the G140L grating. We estimated the proximity sizes 
in the literature. For most of them, their systemic redshifts 
are from the SDSS catalog. The references for their redshifts are listed in 
Tables 3 and 4. One STIS spectrum of quasar PKS1935$-$692 is used. 
\subsection{Measurement of Proximity Zone Size}\label{sub_res}

The study of \he\ proximity profiles is more challenging than that of \h\ at $z\approx 6$ as they are more complex.
We measured a proximity zone from the \he\ \lya\ wavelength to a point where 
the flux drops to below propagated errors as calculated in a bin of approximately 1.5 \AA. 
We used these nominal bin sizes and combined the redshift errors to estimate the 
uncertainties in zone sizes. A difference of 1.5 \AA\ corresponds to a redshift difference of 0.005 for \he\ \lya.
This definition of proximity zones is based on an assumption of zero continuum flux beyond a proximity zone, 
as implied by the high \he\ opacity at $z\gtrsim 2.8$ \citep{shull,syphers2}.  

An exception is the quasar HS1700+6416 ($z=2.751$), where the continuum flux level blueward of  the \he\ \lya\ wavelength 
is known to be nonzero. In Figure 8 of \cite{syphers5}, the \he\ \lya\ absorption region is plotted 
along with a Keck/HIRES spectrum. We estimated the flux level in two regions where
no strong \h\ absorption is present: $2.652 < z < 2.653$ and $2.6545 < z < 2.655$. The average \he\ \lya\
opacity there was estimated as $\tau = 0.95$. Extrapolating this value to $z=2.73$ using a theoretical 
scaling relation 
$\tau_{eff} \propto (1+z)^{3.5}$ \citep{fardal}, the normalized flux level there should be 0.36 ($\tau =1.03$).  
This is approximately the flux level around $z\simeq 2.728$ and 2.739, where no strong \h\ absorption 
features are present, and a potential flux excess near these two redshifts is marginal at best. 
It is likely that the proximity zone in HS1700+6416 ends at the 
absorption feature of $z\simeq 2.744$. A generous estimate would mark the zone down to $z\simeq 2.73$, where 
there is no significant \h\ absorption line.
We therefore list the zone size as $4 \pm 4$ Mpc in Table \ref{tbl-3}, to reflect the two possible 
values of 2 and 6 Mpc and other uncertainties.

All broad proximity profiles are embedded with strong absorption features. At a relatively high
flux, these discrete absorption lines can be easily identified. However, at a low-flux end, they may cut off
the residual flux and cause an underestimate of the proximity zone size.
When an optical spectrum is available, we checked whether a strong \h\ absorption line is present near the 
wavelength that corresponds to the \he\ proximity zone's endpoint. In six cases in Tables \ref{tbl-2} and
\ref{tbl-3}, no such a \h\ absorber is identified within 6 \AA\ of the endpoint's wavelength 
in optical spectra ($dz/(1+z)=0.001$). In the other 11 cases such \h\ absorption features are found near the proximity-zone endpoint, and for the other
seven quasars no optical spectra are available. Therefore the possibility is real that their proximity  
zones may be underestimated. In these 18 cases
we increased the measurement errors by adding a term of 3 Mpc ($dz/(1+z)=0.0035$ at $z=3.5$), approximately the 
nominal width of a strong absorption feature in G140L spectra. The quasar HE2347$-$4342 does not show a proximity zone, and there is no absorption feature in the UVES 
spectrum near the systemic redshift that can account for a potential \he\ LLS.

\subsection{Correlation with Redshift and Luminosity}\label{sub_cor}

Given the sample size, it would be informative to study the evolution of \he\ proximity zones with redshift
and luminosity, which has been reported for \h\ at $z\approx 6$ (\S \ref{sub_comp}). 
We plot the distribution of proximity-zone sizes over redshift and luminosity, respectively, in
Figures \ref{fig-z} and \ref{fig-lum}. The redshift scales in the top and bottom panels of Figure \ref{fig-z} 
are set to reflect the same ratio of cosmic time. Our results 
demonstrate that the proximity effect is common among \he\ quasars, and their proximity-zone sizes 
can be considerably larger than those in the \h\ quasars at $z \approx 6$. The most significant trend seems to
 be at $z>3.3$, where the zone size decreases towards higher redshift. This trend may be understood in terms of 
an increasing \he\ fraction and signals the onset of the intergalactic \he\ reionization.

The two most luminous quasars,
HS1700+6416 and HE2347$-$4342, are located at lower left in the top panel of Figure \ref{fig-lum} and clearly disconnected 
from the rest of the sample. 
Recently a bright quasar at $z=6.3$ was discovered \citep{wu}: with a similar luminosity and a moderate 
zone size, this quasar is also at odds with the model prediction that luminous quasars display a larger proximity zone. 

We ran statistical tests to check potential correlations. The tool is the IRAF astronomical survival analysis 
package, which allows upper or lower limits as inputs \citep{isobe}.
A moderate anti-correlation between zone sizes and quasar redshifts is confirmed: 
the Spearman correlation coefficient $\rho =-0.55$, rejecting a no-correlation hypothesis at 99\% level. 
Proximity-zone sizes do not appear to be correlated with quasar luminosities: 
the Spearman correlation coefficient between them is $0.09$, suggesting a 65\% probability for no correlation. 
Even without the two data points at the high-luminosity end, the correlation is still poor. 

\section{DISCUSSION}\label{sec_disc}

\subsection{Uncertainties in Zone Size}\label{sub_error}

A major source of uncertainties in our measurements of proximity zones is the systemic redshifts 
of quasars. The SDSS quasars have redshift uncertainties on the order of $dz/(1+z) \simeq 
0.002$ \citep{hewett}, which are derived mainly at $z<0.8$ when narrow emission lines are present. 
We verified the redshifts of SDSS quasars in our sample with \oi\
emission in the optical spectra. \oi\ emission is a weak feature and consists of multiple 
components, which carry a statistically weighted average of 1303.49~\AA\ \citep{morton, shull}.
Furthermore, \oi\ emission may be blended with \ion{Si}{2} $\lambda 1305.42$ \citep{vb}.
Assuming a solar abundance, the weighted average for \ion{O}{1}+\ion{Si}{2} is 
1304.22~\AA. We adopted this value with an uncertainty of 1~\AA, which is comparable to
the statistical errors of 5 \AA\ (in the observer's frame) in our fitting to the SDSS spectra. 
Combining these terms and converting into the observer's frame for \he\ \lya, we
estimated uncertainties of approximately 2~\AA\ in the UV spectra of our sample. This term 
corresponds to $dz/(1+z)=0.0015$, 450 \kms\ in velocity dispersion, and 1.2 Mpc in proper 
distance at $z=3.5$. 
We verified the redshifts of SDSS quasars using fits to \oi\ emission when possible.
In a number of cases, \oi\ emission is too weak to be useful in redshift estimates, 
we then used the redshift values from DR7. 
For the non-SDSS quasars 
in Tables \ref{tbl-2} and \ref{tbl-3}, we assumed a redshift error of $\sim 800$ \kms\ in velocity space, unless
quoted explicitly.

Another source of uncertainties in our estimates of the proximity-zone sizes arises 
from their ending points in the low-flux regions.
Our data are mainly moderate-resolution spectra with limited S/N levels. 
For the COS G140L data, this nominal uncertainty is 1.5~\AA, which  corresponds to 
approximately 0.9 Mpc in proper distance at $z=3.5$ and 1.1 Mpc at $z=3.2$.
As discussed in \S \ref{sub_res}, the frequent occurrence of strong absorption features (\h\
column density $\approx 10^{14}$ \cl) adds a term of uncertainty in many cases.

Other rare effects, such as an infalling absorber could affect the redshifts, but are unlikely
to be thick enough to account for the lack of a proximity effect \citep{shull}. 

\subsection{Effect of Intervening Hydrogen Lyman Limit Systems}\label{sub_lls}

Quasar spectra display the signature of numerous IGM components, and density 
fluctuations have a significant impact on the \he\ ionization process. 
LLSs of column density $\gtrsim 10^{17}$~\cl, both in \h\ and low redshift and \he\ at high, 
produce random flux cutoffs.
At $z\simeq 0.5$, a \h\ LLS  may mimic a \he\ absorption edge. The probability of encountering such a system per unit 
redshift is $dn/dz =0.28 (1+z)^{1.19}$ \citep{ribaudo}, which yields a rate that is consistent with other 
results \citep{lsl, esl}. 
Over a wavelength range of 15~\AA\ ($\sim 10$~Mpc in proper distance) near \he\ \lya, there is less than 2\% 
probability of intercepting a \h\ LLS of optical depth $\tau >1$. This value would be even smaller if we 
consider this effect only in the region near the proximity-zone's endpoint.

\subsection{Effect of intervening \he\ Lyman Limit Systems}\label{sub_den}

For \he, the IGM components have considerably higher column densities and a filtering power. 
If the ratio of column densities $\eta = N_{HeII} / N_{HI} \gtrsim 100$, even a moderate \h\ absorber may become
a potential \he\ LLS greatly reduce the number of \he-ionizing photons from 
the quasar, reducing both the growth of a \he\ ionization zone and the subsequent ionization rate. 

A simple estimate \citep{mcquinn} suggests that a system becomes opaque ($\tau > 1$) to the ionizing 
photons at 4 Rydberg if its \h\ column density $N_{HI} > 1.3 \times 10^{16} (50 / \eta)$ \cl. 
Based on the high-resolution data at $z\simeq 2.8$ \citep{gak,zheng04,fechner}, the mean value of $\eta$ is 
around 80, which is derived from a majority of weak components (column density $N_{HI}<3\times 10^{13}$ \cl). 
At $3<z<3.2$, the mean $\eta$ value may be around 200 \citep{syphers6}. However, very high 
$\eta$ values are mostly 
associated weak lines and very difficult to confirm, particularly at $z>3$ when the \he\ opacity is high. 
Evidence suggests that strong components display lower $\eta$ 
values: \cite{fechner} found $\eta \ltsim 15$ at $N_{HI} > 3\times 10^{14}$ \cl, which are supported by 
the data in another bright quasar \cite[$\eta \ltsim 10$ at $N_{HI}> 10^{15}$ \cl,][]{zheng04}. 
Adopting a fiducial value of $\eta = 15$, 
only the systems of \h\ column density  $\gtrsim 4 \times 10^{16}$ \cl\ may filter out the quasar \he-ionizing radiation.
From Table 3 of \cite{kim2013} the distribution can be expressed as 
$d^2n/ (dN_{HI}dX) = 10^{8.64}\ N_{HI}^{-1.54}$. The probability of encountering \h\ systems with $N_{HI} > 4 \times 10^{16}$ \cl\ within 
1.5~\AA\ to the endpoint of a proximity zone (see \S \ref{sub_res}) is estimated as $\sim 0.02$. 

Theoretical models, however, predict high $\eta$ around 100, based on our knowledge of the quasar's 
EUV (extreme UV) continuum, 
and a re-analysis of the average $\eta$ values at $z\approx 2.5$ \citep{mcquinn2014}
does not find evidence for the extremely low $\eta$ values noted in previous studies.
This recent analysis is based on G140L data, while the previous studies used high-resolution FUSE data.
To check the reality of low $\eta$ at $z\gtrsim 3$, we used the published results of QSO0302$-$003 at 
a higher redshift.
\cite{syphers6} used a high-resolution Keck spectrum and COS/G130M spectrum to derive $\eta$ values with a bin size 
of $dz = 0.001$. In the range of $3.03 < z < 3.235$ there are 21 \h\ absorption lines at $N > 10^{14}$~\cl\  
\citep{kim2002}. At $2.76 < z < 2.93$ and $3.25 < z < 3.267$ there are additional eight absorption lines with
restframe equivalent widths greater than 0.3 \AA. From Figures 10-13 of \cite{syphers5} we estimated 29 $\eta$ values: 
twelve of them are lower limits and 17 are measurements. 
For the twelve lower limits, their mean value is $\eta =12$. For the  17 measured $\eta$ values, their 
mean value is 13.

If \he\ LLS are common, no broad proximity profiles should have been observed. We analyzed the 
SDSS spectra of our sample and identified 40 strong \h\ absorption features in the proximity 
zones: nearly all see a significant flux
at their blueward wavelengths. If the $\eta$ values are around 100 for these absorbers, they
should display a sharp flux cutoff. 
Furthermore, in the SDSS spectra of five quasars, no strong \h\ absorbers are found 
near the wavelengths that correspond to a proximity-zone endpoint (Tables \ref{tbl-2} and \ref{tbl-3}). 
The upper limit to the \h\ column density would be  approximately $10^{14}$ \cl, assuming a minimum equivalent 
width of 0.3 \AA\ in the restframe and a nominal velocity dispersion of 30 \kms. 
If an endpoint of the proximity profiles in these quasars is attributed to an assumed \he\ LLS, 
the $\eta$ value may be 8000 or higher, which is not supported by  observations.
We therefore conclude that, while such a possibility cannot be ruled out,  
it is unlikely that proximity profiles are significantly altered by dense \he\ LLS.

\subsection{\he\ Ionized Zone}\label{sub_ion}

The Str\"omgren radius is generally calculated from a balance of the ionizing photon 
rate with the total recombination rate \citep{stromgren,cen,agn,syphers}. 
In a simple model, we assume that the IGM is homogeneous around a quasar, and the 
quasar radiation is isotropic. The total quasar ionizing luminosity in the restframe, 
in units of photon rate, is calculated as
$\dot N_{\gamma} =  4 \pi D_L^2 (1+z)^{-1}\int^{\infty}_{\nu_0} (f_\nu/h\nu) d\nu$ 
where $D_L$ is the bolometric luminosity distance, $f_\nu$ is the flux density in the 
observer's frame and $\nu_0$ is the observed frequency corresponding to 4 Rydberg. 
The Str\"omgren radius is then:
\begin{equation}\label{eq:s}
 R_{S} = \left( \frac{3 \dot N_\gamma}{4 \pi \alpha n_{e} n_{He_{III}}} \right)^{1/3}   \approx 73 \left( \frac{\dot N_\gamma / 10^{57} s^{-1}}{\Delta^2} \right)^{1/3} \left(\frac{1+z}{4}\right)^{-2}\ \ \  {\rm Mpc} \end{equation}
where $\alpha$ is the Case-A recombination coefficient of \heiii\ ions, $n_e$ the electron 
density, and $\Delta$ the local baryon overdensity in units of the mean IGM density,
and $R_{\rm S}$ the proper distance.  The mean density of $1.85 \times 10^{-5}$
 \cc\ at $z = 3.5$ 
is derived from its value at $z=0$ \citep{wmap,syphers}.

The value given in Equation (1) is an upper limit; real ionized zones are considerably smaller for at least 
two reasons. 
First, this hypothetical limiting radius grows faster than the ionization front expands due to the decreasing IGM density
as the Universe expands \citep{shapiro,donahue}. Secondly, for intergalactic helium, 
the recombination time scale is so long, $(\alpha n_e)^{-1} \approx 1$ Gyr, that 
it exceeds the nominal quasar ages. As a result, the IGM around 
quasars would not reach a balance to form a Str\"omgren zone.
If the gas is denser than average ($\Delta > 1$)
or clumpy, and if the quasar age is old, the recombination rate may become comparable 
to the ionization rate.
A proximity zone does not mean that a Str\"omgren sphere has been established,
in the sense of numbers of recombinations balancing numbers of ionizations. This calculation could just indicate that 
the gas within the sphere is over-ionized by the presence of the quasar radiation compared with that in the general IGM. 
The timing criterion for the ionization fraction to stay in ionization balance is only that the age be longer 
than (ionization rate)$^{-1}$ for highly ionized gas (M09). 

Assuming an average IGM density, the opacity is estimated to be on the order of 
$2 \times 10^5\ [(1+z)/6]^{3/2}$ for hydrogen and  
$4 \times 10^3\ [(1+z)/4]^{3/2}$ for \he\ \citep{me,syphers}. 
\cite{syphers3} provide a more accurate formula that can be written for the chosen cosmological parameters:
\begin{equation}\label{eq:tau_gp}
\tau_{{\rm GP, HeII}} = 3.49 \times 10^3\  \Delta\ x_{\rm HeII} 
\left( \frac{Y_P}{0.2486} \right) 
\left( \frac{1+z}{4} \right)^{3/2} \nonumber \\
\left[ \frac{1.0209}{1+0.0209 \cdot (4/(1+z))^3} \right]
\end{equation}

\noindent
where $x_{\rm HeII}$ is the \he\ fraction and $Y_p$ is the helium mass fraction.

In the vicinity of a luminous quasar, the ionizing 
radiation field is enhanced over any metagalactic ionizing background 
radiation, producing a cosmic ``bubble'' that may be observed in the quasar 
spectrum as a proximity effect. This effect should apply to \h\ \citep{murdoch,bdo} as well as \he\ \citep{zheng95,hogan} absorption. 
However, the overall \he\ optical depth could be as high as 
$\approx 4\times 10^3$ if the intergalactic helium is in a form of singly ionized 
state and remains poorly constrained at $z\gtrsim 3.3$.

In the wavelength region 1325-1340 \AA\ ($z=3.362-3.411$) of the spectrum of \obj, the effective 
optical depth is $\tau_{HeII} = 4.10^{+0.43}_{-0.33}$  at 95\% confidence level. 
If we used a wider slit width as the {\tt calcos} 
pipeline default, $\tau = 5.10^{+1.02}_{-0.52}$ at 68\% confidence level, or
$\tau=5.10^{+\infty} _{-0.87}$ at 95\% level. It is therefore possible that the flux extends beyond 12 Mpc 
from the quasar, suggesting that the metagalactic \he-ionizing radiation field is extremely weak but present,
 or the quasar has a \he-ionizing luminosity considerably stronger than the extrapolation from a nominal 
power law, or the ionization level has not yet recovered after an earlier episode of an enhanced quasar
luminosity.

\subsection{Effect of Quasar Age}\label{sub_age}

The proximity profile is subject to the quasar age. A large zone size requires sufficient ionizing photons from the quasar, even after a long period, if 
the surrounding intergalactic helium was not fully ionized, and a large number of photons are needed to ionize it.
For a young quasar, the proximity profile would display a sharp cutoff. Several factors affect the proximity profile: (1) the IGM density and its fluctuations; 
(2) the quasar luminosity; (3) the quasar age; and (4) the recombination process. 
For the case when the quasar age is considerably smaller then the recombination time scale, 
the total number of \he-ionizing photons over a lifetime $t_Q$ is equal to that of 
the ionized helium atoms in the quasar vicinity:
 
\begin{equation}\label{eq:time}
 R_{\rm t_Q} = \left( \frac{3 \dot N_{\gamma} t_Q}{4 \pi n_{He}x_{\rm HeII}} \right)^{1/3} 
\approx 13 \left( \frac{(\dot N_{\gamma} /10^{57} s^{-1})(t_Q/10^7 yr)}
{x_{\rm HeII}\ \Delta}\right)^{1/3} \left(\frac{1+z}{4}\right)^{-1}\ \ \ {\rm Mpc}.
\end{equation} 

We used the proximity profile in \obj\ to model the evolution of intergalactic helium. 
While other cases of 
significant proximity effect exist, the profiles are often shelf-like, possibly due to the duty 
cycle of the quasar or the transverse proximity effect by other quasars \citep{syphers6}. The G140L spectrum, after a reddening correction of $E(B-V)=0.02$, $R_V=3.1$, and the extinction curve of \cite{ext},
displays a threefold increase in flux from 1950 to 1400 \AA, which is 
steeper than that in QSO0302$-$003. Since no LLS break is obvious in this wavelength range, we fitted the
dereddened continuum between 1400 and 1900 \AA\ with a simple power law and a potential partial LLS 
between 2000 and 3000 \AA. The best fit suggests $\beta = -2.09$ ($f_\lambda \propto \lambda^{\beta}$) and a LLS of
$\tau=0.83$ at $\sim 2450$ \AA. If no LLS is assumed, the best-fit power law has $\beta = -2.85$. Both fitting results 
suggest a similar continuum level of $(4.6 \pm 0.5) \times 10^{-16}$ \ergscmA at the \he\ Lyman-limit wavelength 
of $1020$ \AA. 
The estimated total photon rate at 4 Rydberg 
is $ (5.9 \pm 0.6) \times 10^{56}$~sec$^{-1}$. 
As shown in the next section, this hard continuum is not strong enough to produce the observed broad proximity 
profile if the intergalactic helium is not fully ionized.

\subsection{Time Dependence of Ionization}\label{sub_time}

The ionized zones represented by Equations (1) and (3) are extreme cases; the real ionization 
by a quasar is a gradual and slower process, which can be described by an analytical expression 
derived from a simplified differential equation \citep{cen,syphers6} for a fixed IGM temperature.
A more accurate evaluation of the ionization fraction requires
simultaneously solving for the gas temperature and the ionization.
We computed the time dependence of \he\ ionization around a quasar, as outlined in M09. Specifically, we solved the 
time-dependent spherically symmetric set of coupled ionization rate equations for uniformly distributed
hydrogen and helium (Equations 41 and 46 of M09)
on a radial grid, including the photoelectric attenuation by \h,  He~{\sc i} and \he\ of the radiation from 
the quasar. We solved these equations simultaneously with the photoionization heating
and cooling equations as described in M09 (sec.III.B) for a pure hydrogen/helium gas of cosmic
abundances. This calculation used the atomic rates in M09, except for adopting the electron excitation and 
ionization cooling rate of \h\ from \cite{scholz}. Our calculations assumed Case A recombination and 
included Compton cooling off the cosmic microwave background and adiabatic expansion losses in the gas. 
We considered a sequence of quasar activation redshifts corresponding to \heiii-region ages at $z = 3.48$ 
of 10, 20, 30, 50, and 100 Myr.
To match the observed proximity profile, the \he-ionizing continuum must be very strong: 
$L_\nu = 6 \times 10^{31} {\rm erg\ s^{-1}Hz^{-1}}$ at 228 \AA\ in the restframe, which is approximately three times
of the estimated level from observations. It would imply an EUV continuum 
that is even steeper than QSO0302$-$003 \citep{syphers6}. Lower-luminosity models, such as a \he-ionizing 
continuum extrapolated from the flux at \he\ \lya\ wavelength with a nominal power-law index of $-1.7$ \citep{zheng97,telfer} 
were not able to match the size of the ionized region and the flux simultaneously. 
We emphasize that the measured absorption signal is co-temporaneous with the 
quasar observed.
Since the characteristic time to reionize the gas and maintain ionization 
equilibrium is on the order of millions to tens of millions of years, depending on the distance from the 
quasar, the luminosity of the quasar that produced the proximity zone may have been larger than the observed 
value; we have few constraints on quasar variability on million year timescales.

We estimated the black-hole mass as $2.4 \times 10^9$ \Msun, using the \ion{C}{4} line width, the underlying continuum 
flux in the SDSS spectrum and the formula in \cite{vp}. Following the work of \cite{se}, we derived the rest-frame
luminosity near the Lyman limit as $2 \times 10^{47}$ \ergs, which is below the Eddington limit of $3 \times
10^{47}$ \ergs.
A \he-ionizing metagalactic background was not assumed. 
In our simulations of the ionization zone, a hydrogen-ionizing and 
He~{\sc i}-ionizing metagalactic background was turned on at $z = 7$.
Figure \ref{fig-time} shows the \he\ ionization level,
the IGM temperature and the normalized observed \he\ flux at different epochs after the quasar's birth.

Shortly after the quasar's birth, a sharp ionization front expands over time, 
as described by Equation (3). At each radius and each time
interval, the temperature is computed from the energy equation and used to evaluate the atomic rates.
Helium within the sphere is highly ionized, as the ionizing flux is strong and recombination is not effective.
At later epochs, when the ionization front reaches large distances, ionization
balance between the geometrically-diluted quasar ionization flux and
recombinations produces a more gradual rise in the \he\ fraction with distance.
As shown in Figure \ref{fig-time}, a smooth decline of flux may be observed after $\approx 30$ Myr.

The high luminosity assumed in this simulation and that in Equation (3) probably represents an upper limit 
as high-energy 
photons from the quasar need to ionize a large volume of singly ionized helium. This requirement
would be eased if a \he-ionizing background field existed prior to $z \approx 3.5$. As Equation 
(\ref{eq:time}) shows, the ionizing luminosity may be reduced by a factor of three to $6 \times 
10^{56}$ sec$^{-1}$ if the \he\ fraction $x_{\rm HeII}$ is 30\%.
In either case, the ionization front is likely moving nearly at the speed of light, as 
shown in Figure \ref{fig-time}. \cite{white} studied the time-retardation effects on the observed 
proximity-zone size. 
Adapting their equations to \he, the speed-of-light corrections are important to the observed size 
until times large compared with $t_c\approx 100$~Myr for an assumed boosted quasar \he\ ionizing photon 
luminosity of $1.9\times 10^{57}\,{\rm sec}^{-1}$, where $t_c=(3\dot N_\gamma/4\pi n_{\rm HeIII}c^3)^{1/2}$. 
The observed ionization zone sizes then increase slowly with time, as $t^{1/3}$. After an expansion time 
of 50~Myr, the ionization front in Figure \ref{fig-time}
is restricted by the speed of light, but the ionization front velocity slows to 0.5c by 100~Myr. 
According to Equation (3), the age of the observed quasar producing the ionization zone after an expansion 
time of 50~Myr is $t_Q=12$~Myr, which provides a minimal age to the quasar.

If the IGM were highly clumped, the \he\ region would expand
more slowly, as recombinations would slow it down. The high quasar luminosity
used in Figure~\ref{fig-time} is required to match both the position of the \heiii\
ionization front and the \he\ \lya\ flux level behind the front
assuming a homogeneous IGM. If instead the IGM opacity is dominated by
line-blanketing by the \lya\ forest, the quasar luminosity need not
be so high to match the flux level while still matching the position
of the \heiii\ front. This would, however, require a somewhat older
quasar. The model in Figure~\ref{fig-time} demonstrates the extreme assumptions that
must be made for the quasar spectrum to match both the flux level and the
position of the \heiii\ front assuming a homogeneous
IGM. Line-blanketing by the \he\ \lya\ forest is a more plausible
explanation for the flux level \citep{madau94}.

\subsection{Helium Proximity Effect in the Literature}\label{sub_lum}

In principle, a proximity profile should be present in every \he\ quasar spectrum, as a cosmic ``bubble'' 
would exist even without a metagalactic radiation field. At the beginning, ionizing photons do not 
travel very far, as the mean free path $(\sigma_{HeII} n_{HeII})^{-1}$ is less than 0.1 Mpc, where $\sigma_{HeII} = 1.6 \times 10^{-18}$ cm$^{2}$ 
is the photoionization cross section at 4 Rydberg and $n_{HeII} \simeq 2.5 \times 10^{-6}$ cm$^{-3}$ the number density of singly ionized helium at
$z=4$, is less than 0.1 Mpc. High-energy photons from the quasars, however, penetrate considerably deeper into the IGM, raising its
temperature \citep{tittley,mcquinn}. 
These ionized spheres expanded and  eventually overlapped, gradually completing the reionization of intergalactic helium.
While it is anticipated that luminous quasars exhibit a broader proximity profile 
(Equations (1) and (3)), 
it is not the case for the two brightest \he\ quasars at $z \approx 2.8$. 
The UV spectroscopic properties of HS1700+6416 ($z=2.751$, V=16.2) have been extensively studied
\citep{dkz,fechner,syphers5}. Its proximity profile is insignificant, extending at most 6 Mpc (see \S \ref{sub_res}). 
The other well-known quasar, HE2347$-$4342 
\citep[$z=2.887$, V=16.1, ][]{reimers,gak,shull04,zheng04} has no proximity 
effect observed.
The redshifts from \ion{O}{1}~$\lambda$1304 and \ion{O}{3}~$\lambda$5007 agree 
\citep{reimers,syphers2}.
There may be a strong infalling
absorption system observed in the UV spectrum \citep{fechner04,shull}.
\cite{shull} studied this infalling system and
concluded that no satisfactory explanation exists for the absence of a proximity effect around this
luminous quasar other than the possibility that it has only recently turned on (within the last Myr)

\cite{hogan} and \cite{heap} discovered a \he\ proximity zone in the 
quasar QSO0302$-$003 at $z\approx 3.286$ (V=17.5). 
Blueward of the break, a shelf of residual flux corresponding to an opacity
$\tau \approx 0.8 $ extends out about 15 \AA. The recent COS observations
\citep{syphers6} provide details in the proximity zone. The broad proximity 
profile, extending up to 15 Mpc, may be attributed to a line-of-sight effect from 
the quasar itself, or to a transverse effect by another quasar Q0301$-$00 
($z = 3.232$) near the line of sight. 
In the latter case, the age of the ionization zone around Q0301$-$00 should be 
at least 34 Myr.
It is interesting to note that the \he-ionizing continuum is considerably harder
(power-law index $-0.8$) than previous work suggested: broad proximity profiles appear to be 
related to a hard quasar continuum. 
Similarly, in PKS1935$-$692 \citep[$z\simeq 3.18$,][]{anderson}, a proximity shelf of flux extends 
blueward of the \he\ break by at least 20 \AA. This absorption feature displays a strong recovery void at 1246.5 \AA, 
likely produced by the radiation of a foreground source.

\subsection{Comparison with Hydrogen Proximity Effect}\label{sub_comp}

The reionization of intergalactic hydrogen ended by redshift $z\approx 6$, as 
characterized by a sharp increase in opacity at $z>5.7$ \citep{fan}. 
This trend is similar to what is observed for helium 
at $z\approx 3$, and it is instructive to compare these two major cosmic processes.
The number of \h-ionizing photons of a quasar is considerably higher than that of \he-ionizing photons.
On the other hand, the IGM at $z\approx 6$ is denser than that at $z\approx 3.5$. 
The recombination time scale of hydrogen is lower than helium. 
As a result of these factors, our estimate based on Equation (1) suggests that the proximity profile of intergalactic hydrogen at 
$z \approx 6$ is somewhat weaker than that for \he\ at $z\approx 3$. 

\cite{carilli} reported proximity profiles of hydrogen in the spectra of 27 quasars; their ``near zones'' (proximity) lie in the range of $\sim 5-10$ Mpc. They found a significant trend for a decrease
in the near-zone size with increasing redshift, as evidence for 
the evolution of an increasing  neutral fraction of intergalactic hydrogen
toward higher redshifts. They also found that the near-zone sizes increase with the quasar 
UV luminosity, as expected for photoionization dominated by quasar radiation.  
We add eight more data points from recent literature.  One quasar is at the highest known redshift $z \simeq 7.08$ 
\citep{mortlock} and another is luminous \cite[$M = -29.3$, ][]{wu}. These two data 
points carry a significant weight in the correlation analysis. The paper of \cite{ven1} does not list
the near-zone sizes, and we made estimates from their Figure 4.   
A Spearman test over these 35 data points finds a correlation coefficient $\rho = -0.59$ between zone sizes and redshifts, rejecting a no-correlation hypothesis at 99.95\% level. 
Another test between zone sizes and absolute magnitude
finds $\rho =-0.47$, rejecting a no-correlation hypothesis at 99.4\% level. 

The luminous quasar J0100+2802 displays a proximity zone of 8 Mpc. While this is a significant size, it 
does not scale well to luminosity (see the lower panel of Figure 
\ref{fig-lum}). Its normalized zone size, $R_{norm} = R \ 10^{0.4(27+M_{1450})/3}$  
at the restframe wavelength of 1450 \AA, 
is 4 Mpc, below the average at the absolute magnitude $M_{1450}= -27$. 
This result is actually consistent with the two luminous \he\ quasars that they do not show a 
significant proximity profile. Maybe some other factors, such as the density and ionization level in 
the quasar's vicinity or a young quasar age, play an important role. 

\section{CONCLUSION}\label{sub_conc}

We utilize a sample of 24 \he\ quasars to investigate the interplay between these quasars and the 
surrounding IGM. The ionization zone around \he\ quasars is often more prominent than that for \h\ at $z \approx 6$.
The large redshift range for these quasars allows us to gain insight into the IGM reionization 
over a long epoch before intergalactic helium became fully ionized. The proximity-zone sizes decline 
significantly at $z>3.3$, and it is likely that helium reionization started well before $z =3.8$. 

In the quasar SDSS1253+6817, the source flux extends considerably blueward of the \he\ \lya\ 
wavelength, suggesting a quasar age of $\gtrsim 12$ Myr. 
The UV flux rises dramatically from 1950 to 1400 \AA, suggesting an exceptionally hard EUV continuum.  
The $\eta$ value in the proximity zone is lower, consistent with such strong \he-ionizing radiation
from the quasar that produces a broad proximity zone.

The two brightest quasars do not display a significant \he\ proximity profile. While this is 
at odds with model expectations, we notice that a luminous quasar does not display the largest 
\h\ proximity profile either. It is possible that these hyperluminous quasars are young, or they are 
surrounded by a dense IGM. 

It should be stressed that the observed proximity-zone size is not a
direct measurement of the quasar lifetime, as the structure of the
zone is affected by many possible factors, including the IGM density
fluctuations and quasar variability over IGM ionization timescales. 
A quasar with small proximity zone may be considerably younger 
compared with the light-travel time across its proximity zone allowing for retardation effects.
Further understanding of the quasar lifetime awaits improved IGM simulations
that take these factors into account. 

\acknowledgments 
We thank the anonymous referee for many thoughtful and instructive suggestions.

Support for this research was provided by NASA through grant GO-12249 from the Space Telescope Science Institute,
which is operated by the Association of Universities for Research in Astronomy, Inc., under NASA contract NAS 5-26555. 

This work is based on observations made with the NASA/ESA Hubble Space Telescope, 
obtained at the Space Telescope Science Institute, which is operated by the 
Association of Universities for Research in Astronomy, Inc., under NASA contract 
NAS 5-26555 and observations obtained with the Apache Point Observatory 3.5-meter 
telescope, which is owned and operated by the Astrophysical Research Consortium.

Funding for the SDSS and SDSS-II has been provided by the Alfred P. Sloan Foundation, 
the Participating Institutions, the National Science Foundation, the U.S. Department 
of Energy, the National Aeronautics and Space Administration, the Japanese 
Monbukagakusho, the Max Planck Society, and the Higher Education Funding Council for 
England. The SDSS Web Site is http://www.sdss.org/.

This research has made use of the Keck Observatory Archive (KOA), which is operated by the W. M. Keck 
Observatory and the NASA Exoplanet Science Institute (NExScI), 
and the data obtained from the ESO Science Archive Facility under request number 154374.

\clearpage  

\centerline{\bf APPENDIX A}

\centerline{\bf ACS PRSIM SPECTRA}

Approximately half of all the known \he\ quasars were discovered in three ACS 
prism surveys \citep [GO 10907, 11215, 11982: PI Anderson; ][]{syphers0,syphers1}. 
The ACS PR130L prism offers a low spectral resolution, $R \approx 170-380$ between 
1360 and 1250 \AA. 
While the potential merit of these prism spectra is limited, their
sample size is significant. It is therefore interesting to explore the proximity profiles in the ``other half'' 
of the \he\ quasar sample.
We estimated the proximity profiles in the published prism spectra of 15 quasars at $z<3.6$.
The spectra of six other quasars at higher redshifts were not used because the spectral 
resolution degrades significantly redward of 1400 \AA.

To understand the effect of spectral deconvolution, 
we simulated thousands of spectra with different proximity profiles ($0 - 25$ \AA)
at $z=3.5$ and a pixel scale of 0.1 \AA, then convolved with a Gaussian kernel appropriate to 
the wavelength-dependent ACS resolution. The input spectra were of simple constant flux, with a
proximity profile that starts at 1367~\AA\ and declines linearly towards zero at the end of the proximity zone.
The flux at shorter wavelengths of the proximity zone was set to zero.
The noise level was set for an exposure time of 
4500 sec, which is typical for these prism observations. As the S/N level in prism spectra is
quite high ($> 10$), our simulation results are not sensitive to noise. 
Each of these simulated spectra was smoothed 
to a resolution of the PR130L prism and then binned with small wavelength offsets. 
These offsets reflect the possibility that a spectrum may shift within a prism-data 
pixel and subsequently affect the deconvolution result.  These simulated prism
data were deconvolved using the IDL task {\tt MAX\_LIKELIHOOD} with a point-spread-function 
kernel of two pixels and then compared with the input data.

Our measurements of proximity zones in approximately 90\% of the simulated spectra were 
within an error of one prism pixel ($ \sim4$~\AA), and the others recovered within two pixels.  
A conservative estimate for the errors in prism data as 
7 \AA\ ($\simeq 5$ Mpc at $z = 3.5$), which are considerably higher.

We compared the COS/G140L spectra of quasars SDSS1253+6817 and SDSS2346$-$0016 with their counterparts in 
the ACS/PR130L prism data. 
As shown in Figure \ref{fig-prm}, prism spectra display an extended wing at a level $<10\%$
of the unattenuated flux. This may be due to the extended wing of the instrumental
line-spread function or potential misalignment of a few individual prism images.
We measured the proximity zone to a flux level of 10\% level. 
The proximity profiles in the prism data are consistent with that in the COS/G140L data:
\obj\ displays a broad profile and SDSS2346$-$0016 does not.

Most of the ACS prism data yielded no detection of proximity profiles ($\lesssim 5$~Mpc). 
In Table \ref{tbl-4} we list three quasars whose proximity profiles in the ACS prism spectra
are larger than 7 \AA.  They are at relatively high redshifts, and the data suggest that the decline
of proximity-zone size may start at $z \simeq 3.5$.

\bigskip

\centerline{\bf APPENDIX B}

\centerline{\bf PROXIMITY EFFECT in SDSS DATA}

The study of the proximity effect has been based on high-resolution quasar spectra \citep{murdoch,bdo, dallaglio2}, 
where the numbers of forest lines are compared in the quasar's vicinity and at large distances. 
The effect of enhanced radiation in a quasar's vicinity may be seen even at a low spectra resolution.
\cite{dallaglio} used a flux transmission technique on the VLT/FORS2 spectra ($R\sim 800$) of 17 bright 
quasars at $2.7 < z < 4.1$ to measure the effective optical depth along the lines of sight and detect a 
proximity effect. 

The effective optical depth accumulated from most IGM components is estimated as 
\begin{equation}\label{eq:tau}  \tau = 0.0021 (1+z)^{3.7} \end{equation} \citep{meiksin06, kirkman, da}. 
Approximately half of this term is from 
those at column density $N > 10^{14}$ \cl,
most of which cases can be identified at the SDSS spectral resolution and S/N 
(\S \ref{sub_res}). From a  power-law distribution of 
$dn/dN_{HI} \propto N_{HI}^{-1.6}$ where $N$ is the column density and $n$ the number of 
individual absorbers \citep{tytler,janknecht}, we estimated approximately six strong absorbers over a range of 
10 Mpc ($\sim 66$ \AA\ on the SDSS wavelength scale). These lines affect roughly 60\% of the pixels in
this bin, assuming that each line affects five SDSS spectral pixels. The fluxes in other pixels in this bin 
are believed to bear the signature of weak IGM components.
Since these lines are unsaturated, their intensity is sensitive to the proximity effect. By measuring the 
flux decrements in these pixels, we tried to detect the signature of proximity zones.

We generated a set of absorption templates $\tau = 0.0021 (1+z)^{3.7}/(1+(R_p/r)^2)$, 
where r is the distance to the quasar and $R_p$ the proximity size. The range of 
$R_p$ is 5$-$35 Mpc. The fitting task {\tt specfit} allows such a ``userabs'' component in the form of opacity vs.
wavelength. We ran {\tt specfit} on SDSS spectra with the following components:
an underlying power law, Gaussian emission components of \lya\ (narrow + broad), \ion{N}{5}, \ion{O}{1}, \ion{Si}{4}, 
\ion{C}{4}, \ion{He}{2}, and a set of absorption
templates. The fitting windows consists of two parts at the following restframe wavelengths: 
(1) 1216$-$1680 \AA, 
which cover the red wing of \lya\ and five other emission lines (\S \ref{sub_cos}) and
(2) a number of small windows between 1200 and 1216 \AA, where no strong forest line is detected. 

We ran tests on the SDSS spectra two quasars with different \he\ proximity zones.
For SDSS1253+6817, our fitting results suggest a symmetrical \lya\ profile, implying a strong proximity profile.
The average opacity $\tau_{eff} = 0.03 \pm 0.03$ over a bin $20-80$ \AA\ blueward of the 
\lya\ wavelength. The calculation includes 22 pixels that are not in absorption windows.
For SDSSJ1711+6052, the fitting result suggests that the blue \lya\ wing is weaker than the red wing, 
as expected from IGM absorption.
The average optical depth in a similar wavelength range is $\tau_{eff} = 0.46 \pm 0.33$.
The S/N in the second quasar is $\simeq 7$ per pixel in the proximity zone, and the errors 
in our measurements are dominated by pixel-to-pixel variations.
A higher average optical depth with noticeable pixel-to-pixel fluctuations is consistent with a 
significant structure in the forest lines and a weak proximity effect. 
These fitting results suggest that the signature of a broad proximity zone may be detected in SDSS spectra.

\clearpage
{
\begin{deluxetable}{lcrccc}
\rotate
\tablecaption{Summary of Observations
\label{tbl-1}}
\tablewidth{0pt}
\footnotesize
\tighttable
\tablehead{
\colhead{Quasar} &
\colhead{R.A.} &
\colhead{Decl.} &
\colhead{Observation} & 
\colhead{Exposure Time} &
\colhead{Grating Central}\\
\colhead{} &
\multicolumn{2}{c}{(J2000)} & 
\colhead{} &
\colhead{(sec)} &
\colhead{Wavelength (\AA)}
}
\startdata
\hline 
\multicolumn{6}{c}{Data of GO 12249}\\ 
\hline
SDSS1253+6817  & 12 53 53.715 & ~68 17 14.20  & 2011 May& 14095&  1105/1280  \\ SDSS2346$-$0016  & 23 46 25.662 & $-$00 16 00.47  &2010 Nov., Dec.& 20737&  1105/1280 \\ SDSS1711+6052  & 17 11 34.412 & ~60 52 40.39  & 2011 Apr., Oct.& 23950&  1105 \\ SDSS1319+5202  & 13 19 14.205 & ~52 02 00.11   & 2011 May& 26643&  1105     \\  \hline 
\multicolumn{6}{c}{Archival Data of GO 11742, 13013}\\ 
\hline
SDSS1237+0126  & 12 37 48.993 & ~01 26 06.90 & 2010 Jun.& 6212&  1105  \\ QSO2149$-$0859  & 21 49 27.770 & $-$08 59 03.61 & 2013 Apr.& 7561&  1105  \\ SDSS0936+2927  & 09 36 43.511  & ~29 27 13.60  &2011 Jan.& 4739&  1105 \\ QSO1630+0435   & 16 30 56.340 & ~04 35 59.42 & 2013 Apr. & 7908  & 1105 \\ 
QSO2157+2330   & 21 57 43.630 & ~23 30 37.34 & 2013 Jul. & 8074  & 1105 \\ 
QSO0233$-$0149 & 02 33 06.010 & $-$01 49 50.58 & 2013 Aug.& 2021 & 1105\\ QSO0916+2405   & 21 57 43.630 & ~23 30 37.34 & 2013 Dec. & 3075  & 1105 \\ 
\enddata
\end{deluxetable}
}

\clearpage

{
\begin{deluxetable}{lcccccc}
\rotate
\tablecaption{Proximity Zone Measurement\label{tbl-2}}
\tablewidth{0pt}
\footnotesize
\tighttable
\tablehead{
\colhead{Quasar} &
\multicolumn{2}{c}{Redshift\tablenotemark{a}} & 
\colhead{Observed Continuum Flux $f_{\lambda 0} (304)$}&
\colhead{Magnitude} & 
\multicolumn{2}{c}{Proximity Zone Size}  
\\
\colhead{} &
\colhead{This work} & 
\colhead{SDSS} & 
\colhead{$10^{-17}$ \ergscmA} &
\colhead{$M_{1450}$} & 
\colhead{\AA} &
\colhead{Mpc}
}
\startdata
\hline 
\multicolumn{7}{c}{Data of GO 12249}\\ 
\hline
SDSS1253+6817\tablenotemark{b}  & $3.476 \pm 0.004$ & 3.4727 &19& $-27.3$ & 20 & 12   \\ SDSS2346$-$0016\tablenotemark{c}  & $3.511 \pm 0.003$ & 3.4895 &19& $-28.0$& 6 & 3.5 \\ SDSS1711+6052\tablenotemark{c}  & $3.823 \pm 0.007$ & 3.8269 & 8 & $-26.6$ & 5  &2.5    \\ SDSS1319+5202\tablenotemark{c}  & $3.93 \pm 0.01 $ & 3.8991 &2& $-28.2$ & $10$ &4.7    \\ \hline 
\multicolumn{7}{c}{Archival Data of GO 11742, 13013}\\ 
\hline
SDSS0936+2927\tablenotemark{b}  & $2.9253 \pm 0.004 $ & 2.9239 & 15 & $-27.4$ & 20 & 16  \\ QSO2157+2330   &  \multicolumn{2}{c}{3.142\tablenotemark{d}} & 15 & $-27.7$ & 26 & 19 \\ 
SDSS1237+0126\tablenotemark{b}  & \multicolumn{2}{c}{3.1448\tablenotemark{d}}  & 15 & $-26.8$ & 11 & 8 \\ QSO2149$-$0859 	& \multicolumn{2}{c}{3.259\tablenotemark{e}} & 6 & $-26.9$ & 4 & 2.7 \\ QSO0233$-$0149 & \multicolumn{2}{c}{3.314\tablenotemark{e}} & 11 & $-27.2$ & 11 & 7.2 \\ QSO0916+2405 & \multicolumn{2}{c}{3.440\tablenotemark{e}} & 18 & $-27.0$ & 7 & 4 \\
QSO1630+0435 	& \multicolumn{2}{c}{3.788\tablenotemark{e}} & 23 & $-28.3$ & 10 & 5.1 \\ \enddata
\tablenotetext{a}{The nominal redshift uncertainty is 0.002 for SDSS spectra.}
\tablenotetext{b}{Optical spectra checked. Strong \h\ absorption found near the end of the proximity zone.}
\tablenotetext{c}{Optical spectra checked. No strong \h\ absorption found near the end of the proximity zone.}
\tablenotetext{d}{\oi\ emission is contaminated by absorption. The redshift value is from \cite{worseck14}.}

\tablenotetext{e}{http://www.stsci.edu/hst/phase2-public/13013.pro}
\end{deluxetable}
}

\clearpage

{
\begin{deluxetable}{llcccl}
\tablecaption{\he\ Measurement Based on Published COS and STIS Data\label{tbl-3}}
\tablewidth{0pt}
\footnotesize
\tighttable
\tablehead{
\colhead{Quasar} &
\colhead{Redshift\tablenotemark{a}} &
\colhead{Magnitude} & 
\multicolumn{2}{c}{Proximity Zone Size} & \colhead{Reference} \\
\colhead{} &
\colhead{} & 
\colhead{$M_{1450}$} &
\colhead{(\AA)} &
\colhead{(Mpc)} &
\colhead{}
}
\startdata
HS1700+6416\tablenotemark{b}& $2.751 \pm 0.003$ &  $-29.4$ & 4 & 3.7 & \cite{syphers5} \\ 
Q1216+1656      & 2.818~ & $-27.3$ & 15 & 13 & \cite{syphers4}  \\ 
HS1024+1849\tablenotemark{b}     & 2.8475 & $-27.0$  & 21 & 18 & \cite{syphers4}  \\ 
4C57.27         & 2.858~ & $-27.1$ & 11 & 9 & \cite{syphers4}  \\ 
HE2347$-$4342\tablenotemark{c}& $2.885 \pm 0.005$  & $-29.3$ & 0 & 0 & \cite{shull} \\ 
SDSS1508+1654\tablenotemark{b}   & 3.1716 & $-27.3 $ & 18 & 12 & \cite{syphers4}  \\ 
PKS1935$-$692\tablenotemark{b}	& 3.185~ &	 $-28.3$  & 25 & 17 & \cite{anderson}  \\ 
SDSS0856+1234\tablenotemark{b}   & 3.1948 & $-27.0 $ & 20 & 14 & \cite{syphers4}  \\ 
SDSS0955+432 \tablenotemark{b}  & 3.2388 &$-26.2 $& 18 & 12 & \cite{syphers4}  \\ 
QSO0302$-$003\tablenotemark{b}   & 3.2860 & $-28.3$ & 17 & 11 &\cite{syphers6} \\ 
SDSS0915+4756\tablenotemark{c}   & 3.3369 & $-27.9$ & 6 & 3.9 & \cite{syphers4}  \\ 
SDSS2345+1512\tablenotemark{b}   & 3.5880 & $-26.6$ & 5 & 2.8 & \cite{syphers4}  \\ 
SDSS2257+0016\tablenotemark{c}   & 3.7721 &$-27.2$ & 5 & 2.6 & \cite{syphers4}  \\ 
\enddata
\tablenotetext{a}{From the SDSS DR7 quasar catalog \citep{dr7}, except: HS1700+6416 \citep{trainor}; HE2347$-$4342 \citep{reimers}; 
QSO0302$-$003 \citep{syphers6}; PKS1935$-$692 \citep{anderson}; Q1216+1656, 4C57.27 \citep{veron}. The nominal redshift uncertainty is 0.002 for SDSS quasars.}
\tablenotetext{b}{Optical spectra checked. Strong \h\ absorption found near the end of the proximity zone.
For HS1700+6416, this line is present at $z=2.744$.}
\tablenotetext{c}{Optical spectra checked. No strong \h\ absorption found near the end of the proximity zone.}
\end{deluxetable}
}

\clearpage

{
\begin{deluxetable}{llcccl}
\tablecaption{\he\ Measurement Based on Published ACS Prism Data\label{tbl-4}}
\tablewidth{0pt}
\footnotesize
\tighttable
\tablehead{
\colhead{Quasar} &
\colhead{Redshift\tablenotemark{a}} &
\colhead{Magnitude} & 
\multicolumn{2}{c}{Proximity Zone Size} & \colhead{Reference} \\
\colhead{} &
\colhead{} & 
\colhead{$M_{1450}$} &
\colhead{(\AA)} &
\colhead{(Mpc)} &
\colhead{}
}
\startdata
SDSS1042+5129   & 3.3864  & $-27.1$ & 26 & 17 &  \cite{syphers3} \\ SDSS1007+4723 	& 3.4084 & $-26.2$ & 9 & 6 & \cite{syphers1} \\ SDSS1442+0920 	& 3.5286 & $-28.3$ & 13 & 8 &  \cite{syphers1} \\ \enddata
\tablenotetext{a}{From the SDSS DR7 quasar catalog \citep{dr7}. The nominal redshift uncertainty is 0.002}
\end{deluxetable}
}

\clearpage

\centerline{\bf Figure Captions}
\bigskip

\figcaption{
\HST/COS G140L spectra of four quasars. The data are binned by six pixels ($\approx 0.5$~\AA, 
0.85 resolution elements).
The green dotted curves are the errors, and the red dashed lines mark the wavelengths 
of \he\ \lya. Vertical bars mark the wavelength of
strong absorption lines identified from the SDSS spectrum counterparts, and 
arrows mark the edge of proximity zones.
}

\figcaption{Near-infrared spectrum of SDSS2346$-$0016, 
taken with the ARC 3.5-m telescope and TripleSpec instrument. Gaps in 
wavelengths are due to the removal of poor data in the bands of 
severe atmospheric absorption. The inset displays the wavelength region 
for redshifted \ion{Mg}{2} emission. We used it to determine 
the systemic redshift of $z=3.511$.
}

\figcaption{
\HST/COS G140L spectra of seven other quasars in the \HST\ 
archive, plotted in the quasar's restframe.
The data are binned by six pixels ($\approx 0.5$ \AA). The green dotted curves are the errors, 
and the red dashed lines mark the wavelengths of \he\ \lya. 
The Earth symbols mark the wavelengths of the strong geocoronal lines 
around 1216 and 1302 \AA, and a weak one at 1356 \AA.
Vertical bars mark the wavelength of the
strong absorption lines identified from the SDSS spectrum counterparts, and 
arrows mark the edge of proximity zones.
}

\figcaption{Normalized spectra of quasar \obj. 
The lower panel is the \he\ \lya\ region in the observer's wavelength frame (COS/G140L), and the upper panel
is the aligned \h\ counterparts (SDSS). The proper distance scale in the top label is appropriate for both 
panels. The green shaded region marks the wavelength range for the \he\
\lya\ break, given the redshift uncertainty ($z = 3.472 - 3.480$). The magenta arrow marks the
wavelength for the ending point of the estimated proximity zone. The blue lines mark respective baselines 
and data errors. The red bars mark the strong absorption features identified in the SDSS spectrum.
}

\figcaption{Size of proximity zone vs. redshift. 
In the upper panel, red boxes are COS data, GO 12249;
magenta boxes: archival data of GO 11742 and 13013; 
green boxes: COS data from literature; 
the cyan star: STIS data of PKS1935$-$692.
The downward arrow is for $1\sigma$ limit with non-detection. The lower panel is for \h. Black crosses are from Table 1 of \cite{carilli}; 
green circles: \cite{ven1,ven2}; and 
red triangles: \cite{mortlock, wu}. 
The redshift ranges in the two panels are scaled to represent the same ratio 
(1.0:0.65) of cosmic ages from the Big Bang. 
}

\figcaption{Size of proximity zone vs. absolute magnitude.  
The symbols are the same as Figure \ref{fig-z}. Yellow curves represent model predictions 
of $R_s \propto L^{1/3}$. The very luminous quasars ($M < -29$) show proximity-zone sizes that are
considerably smaller than anticipated.}

\figcaption{Simulated \he\ ionization structure, assuming a homogeneous IGM. 
The lower panel: log(\he\ fraction), middle panel: gas temperature in units of $10^3$ K, and upper panel: 
normalized flux. 
The quasar \he\ ionization rate has been boosted to $2\times 10^{57}\,{\rm sec}^{-1}$. This extreme case 
assumes the quasar has been observed in a relatively quiescent state. The ionization levels assume the 
boosted rate, as it would take about 9~Myr for ionization equilibrium to be established at the lower 
rate at the ionization front.
Five curves represent an expansion time sequence of 10 (cyan), 20 (blue), 30 (green), 50 (magenta), 
and 100 (red) Myr after the quasar turns on. The age of the quasar producing the ionized zone will generally 
be smaller (see text). The proximity profile in \obj\ is overplotted.}

\figcaption{Comparison of ACS/PR130L and COS/G140L spectra of two quasars. In the top panels, 
the green histograms are the \HST\ data, and black histograms the deconvolved data. The proximity 
zones are marked between the respective magenta arrow and red vertical line, which is at the wavelength 
of \he\ lya. The zero-flux baselines are marked by blue dashed lines. 
}

\clearpage 
\setcounter{figure}{0}
\begin{figure}
\plotone{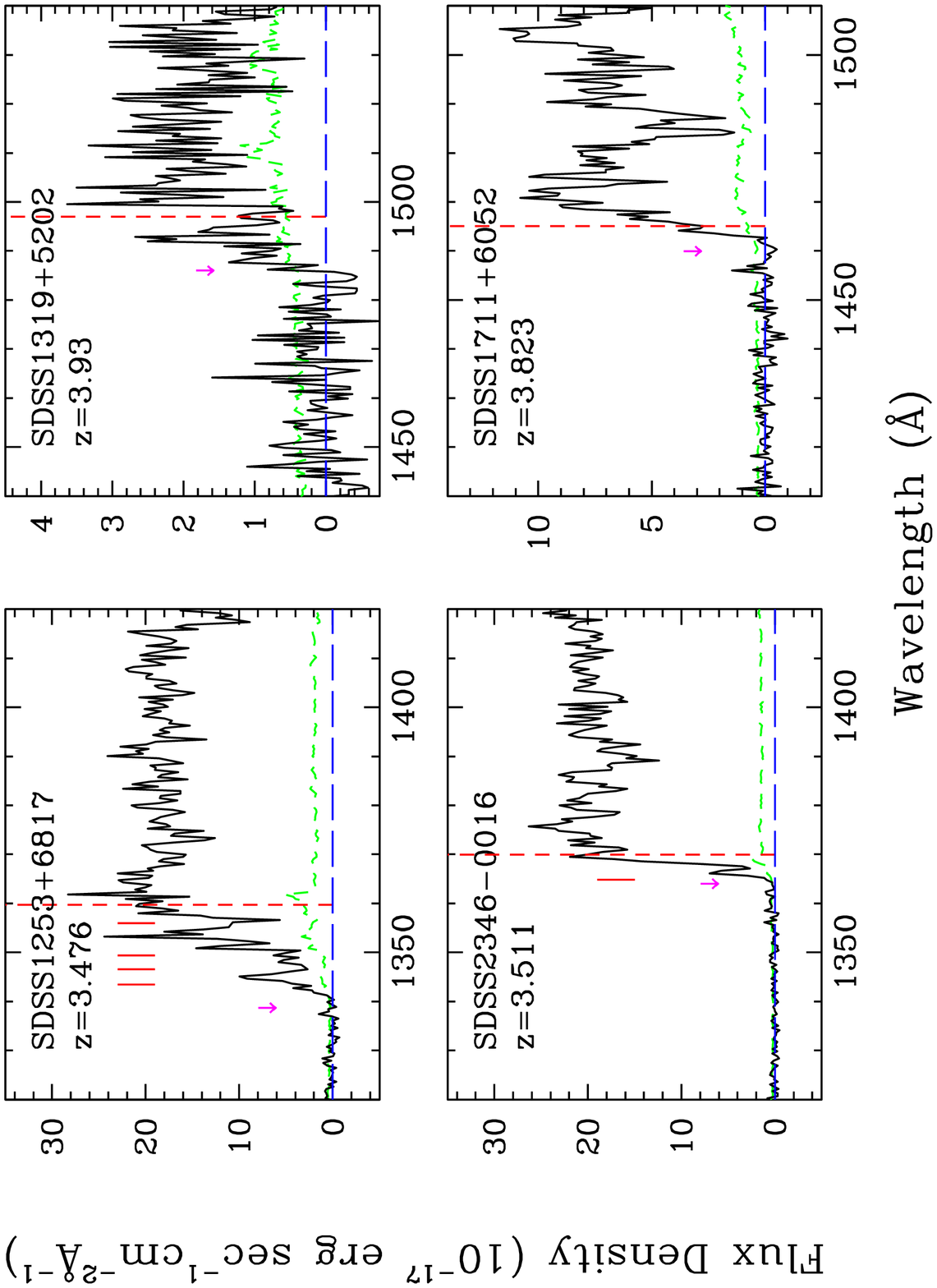}
\caption{
\label{fig-cos}}
\end{figure} 

\begin{figure}
\plotone{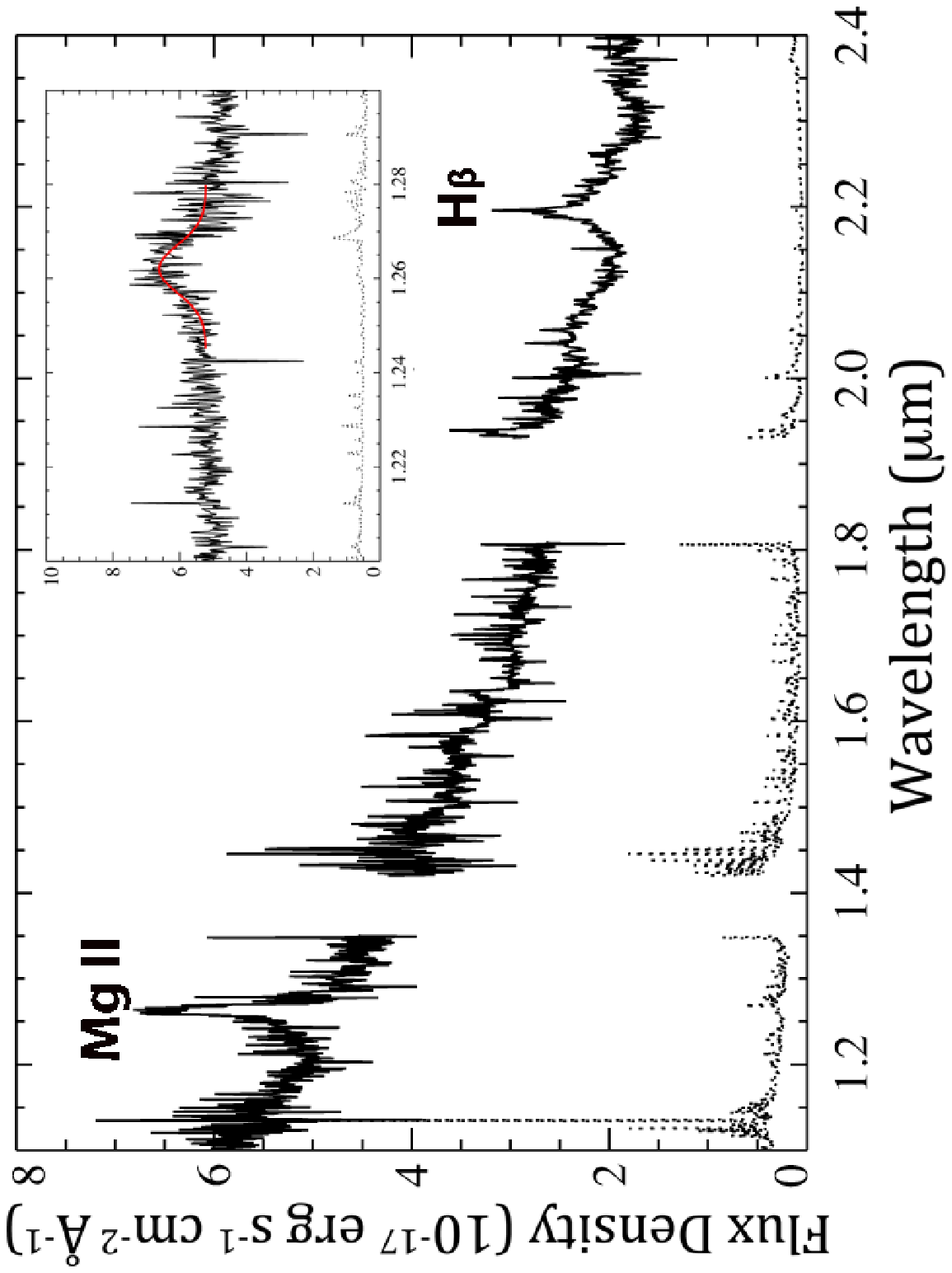}
\figcaption{
\label{fig-nir}}
\end{figure} 

\begin{figure}
\plotone{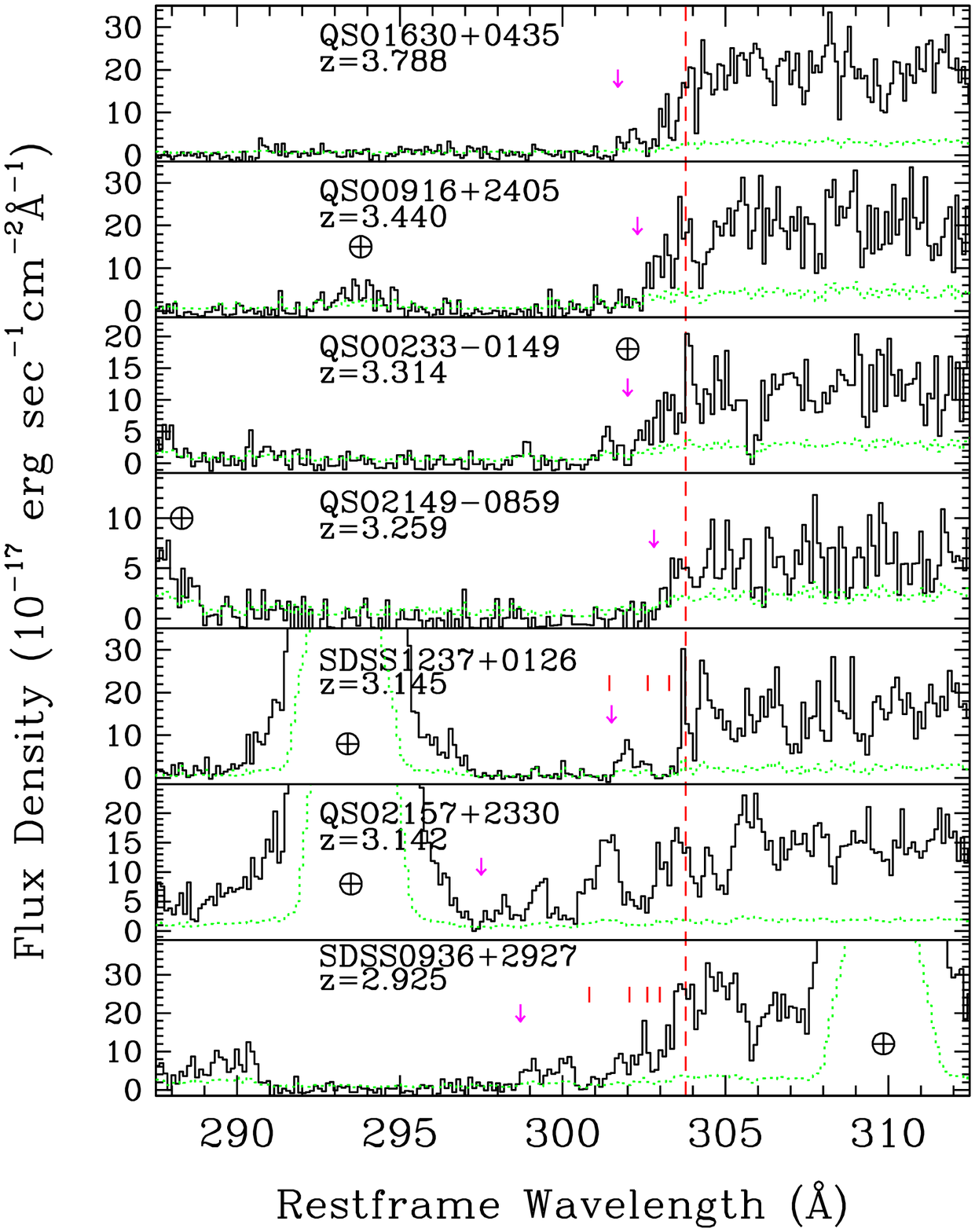}
\figcaption{
\label{fig-cos2}}
\end{figure} 
\clearpage 

\begin{figure}
\plotone{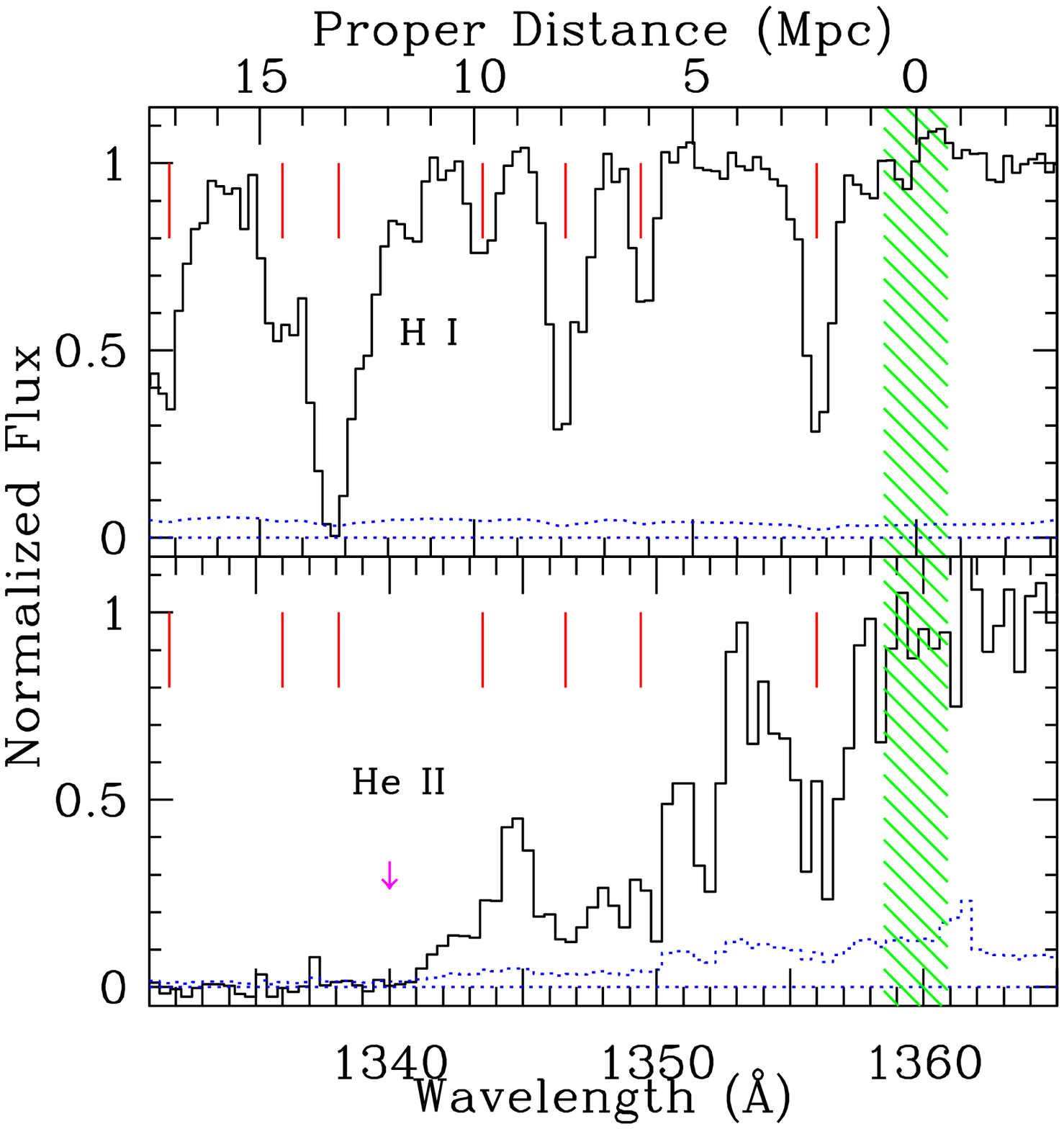}
\caption{\label{fig-fit}}
\end{figure} 
\clearpage 

\begin{figure}
\plotone{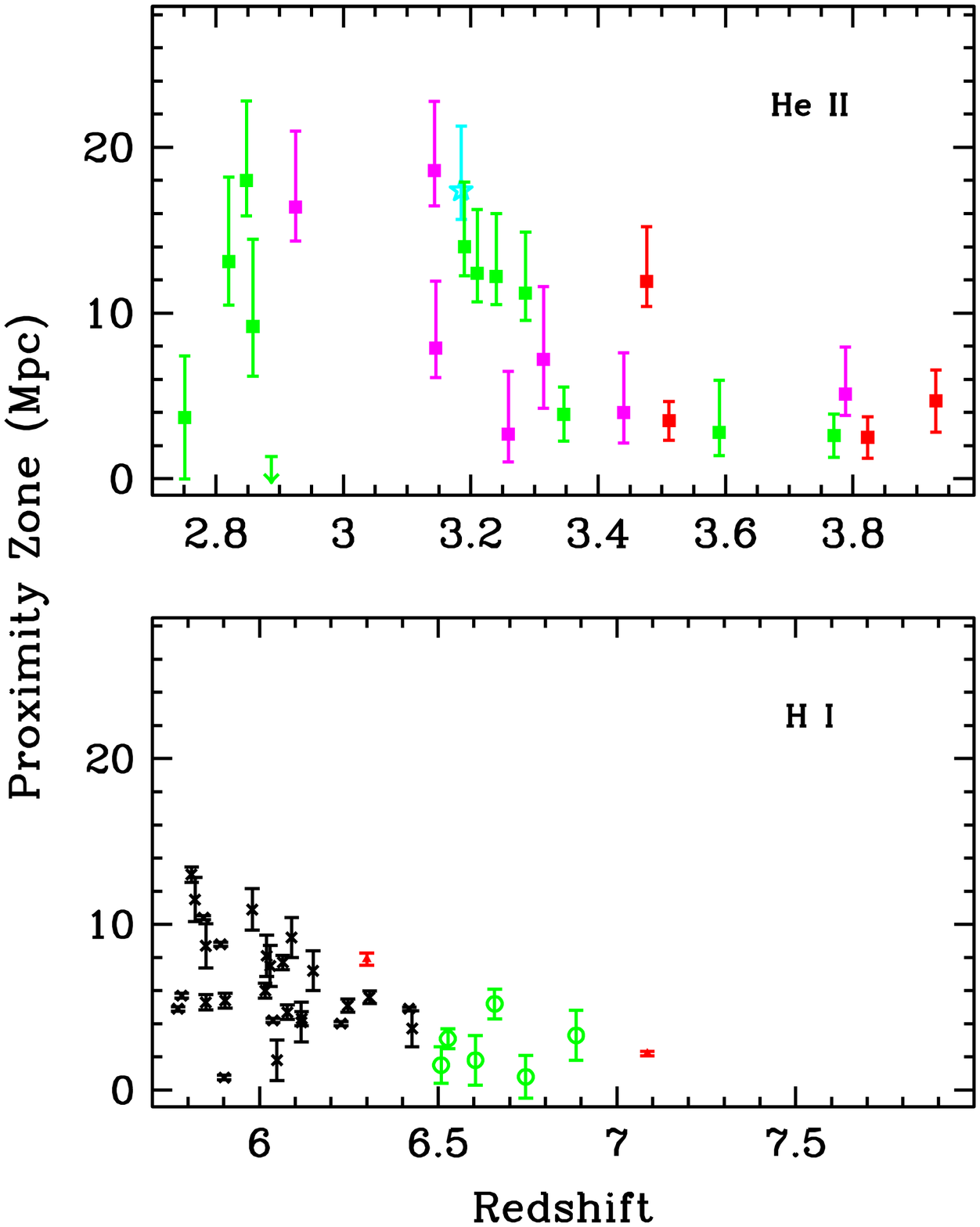}
\caption{\label{fig-z}}
\end{figure} 
\clearpage 

\begin{figure}
\plotone{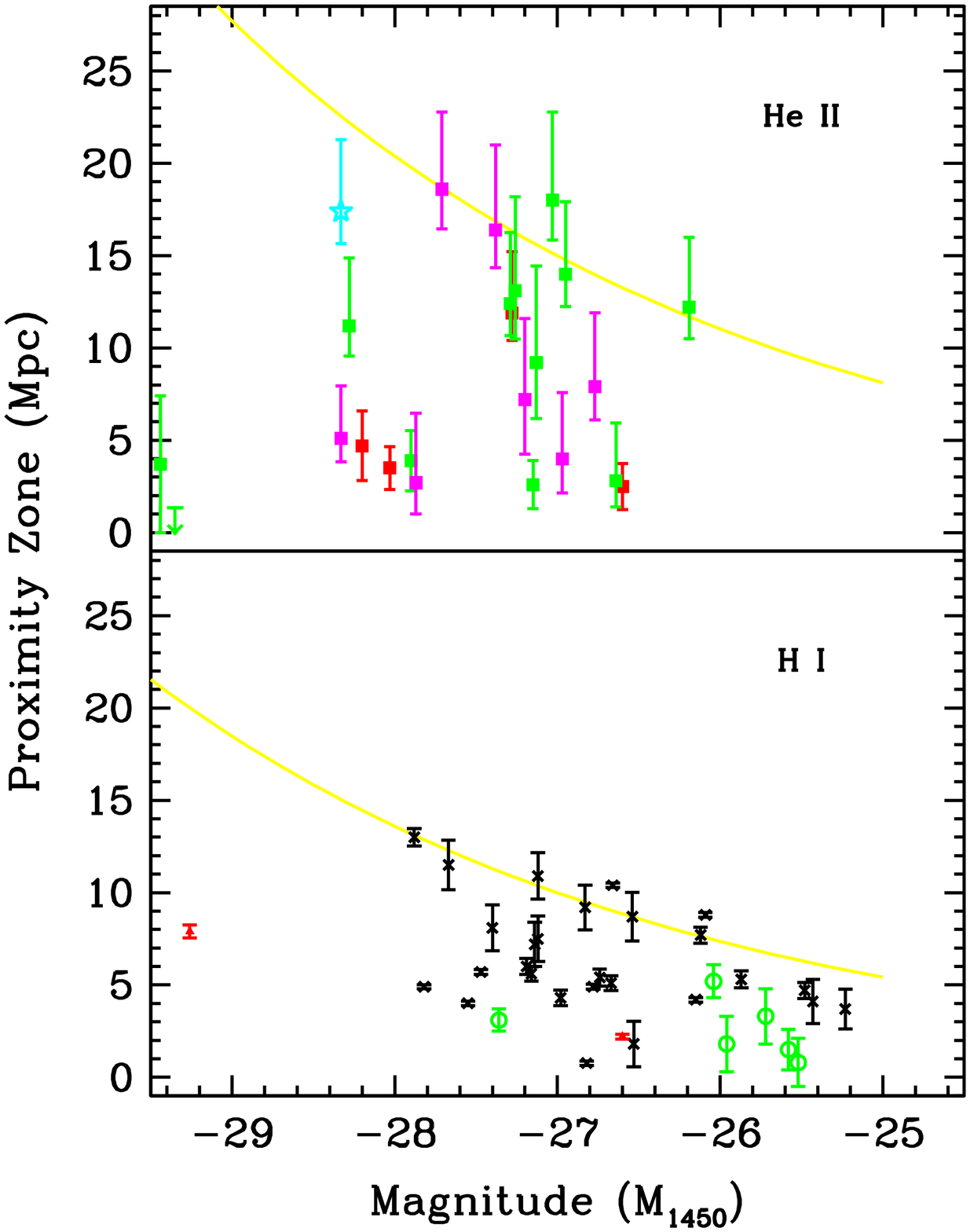}
\caption{\label{fig-lum}}
\end{figure} 

\clearpage 

\begin{figure}
\plotone{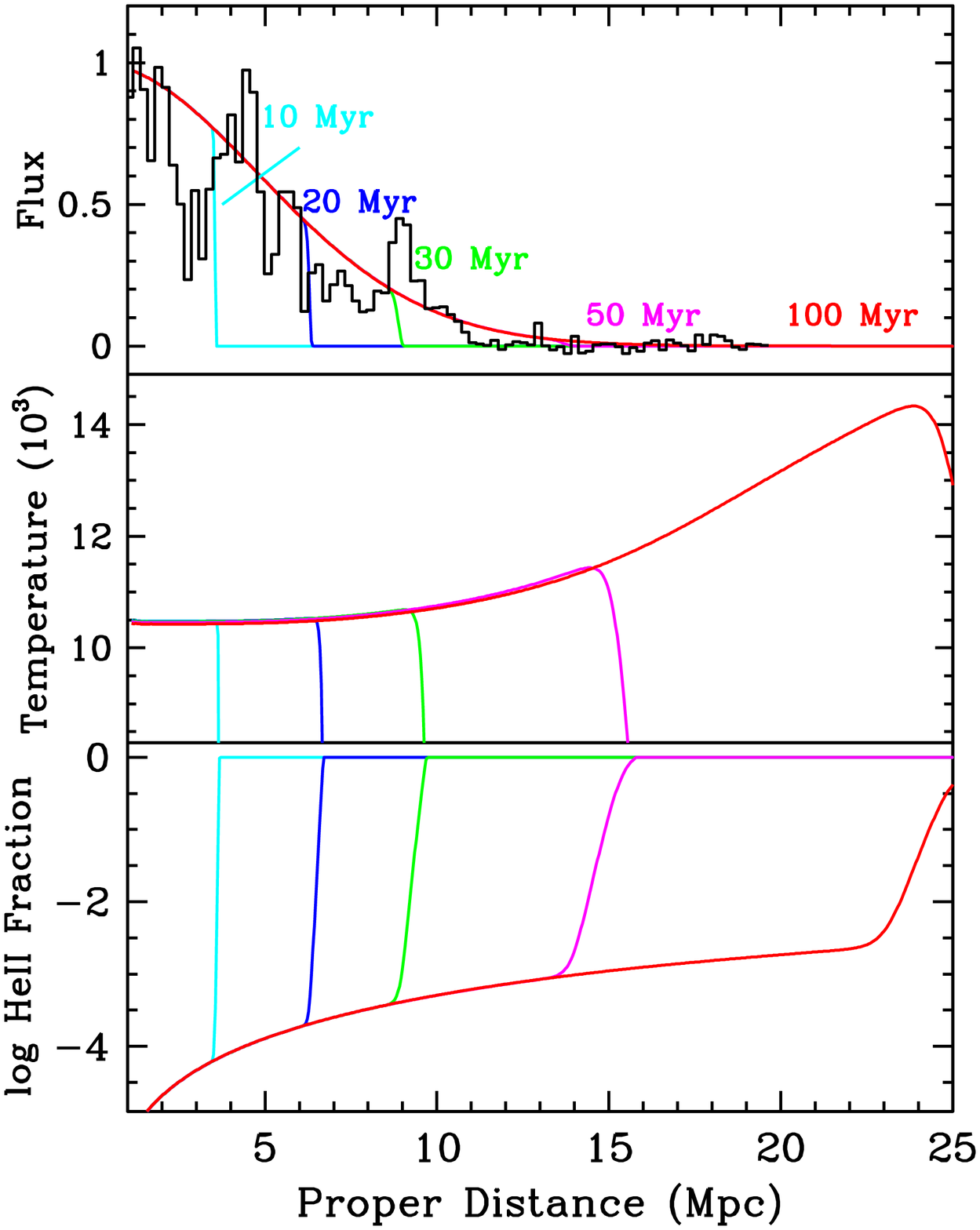}
\figcaption{
\label{fig-time}}
\end{figure} 

\clearpage

\clearpage

\begin{figure}
\plotone{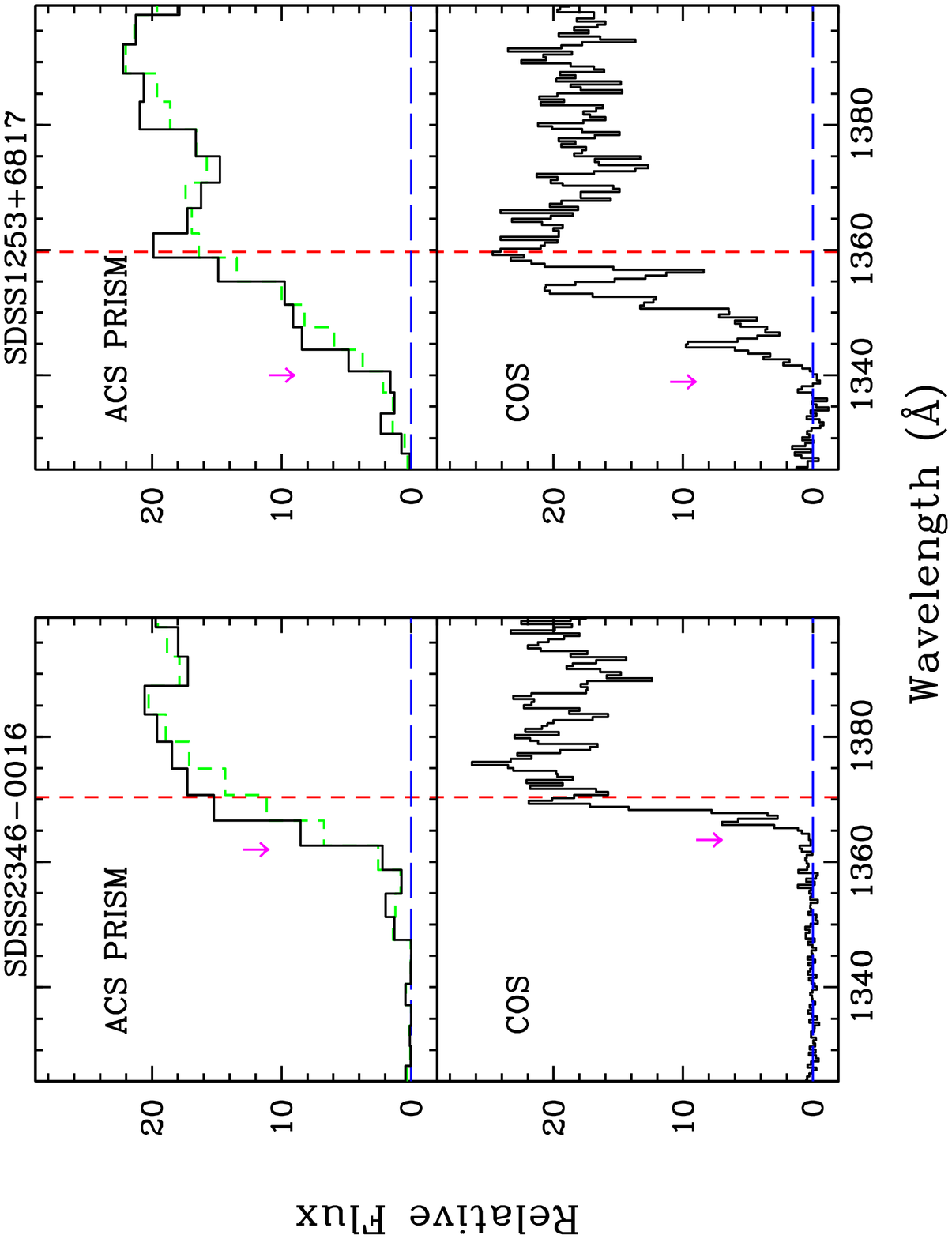}
\figcaption{
\label{fig-prm}}
\end{figure} 

\clearpage 


\begin{thebibliography}{}
\bibitem[Anderson et al.(1999)]{anderson} Anderson, S. F., Hogan, C. J., 
  Williams, B. F. \& Carswell, R. F. 1999, \aj, 117, 56
\bibitem[Bajtlik, Duncan \& Ostriker(1988)]{bdo} Bajtlik, S., Duncan, R. C. \& Ostriker, J. P. 1988, \apj, 327, 570 
\bibitem[Becker et al.(2013)]{da} Becker, G. D., Hewett, P. C., Worseck, G., \& Prochaska, J. X. 2013, MNRAS, 430, 2067
\bibitem[Becker et al. (2001)]{becker} Becker, R. H., Fan, X., White, R. L. et al. 2001, \aj, 122, 2850
\bibitem[Carilli et al.  (2010)]{carilli} Carilli, C. L., Wang, R., Fan, X. et al.  2010, \apj, 714, 834
\bibitem[Carswell et al.  (1987)]{carswell} Carswell, R. F., Webb, J. K., Baldwin, J. A. \& Atwood, B. 1987, \apj, 319, 709
\bibitem[Cen \& Haiman(2000)]{cen} Cen, R. \& Haiman, Z. 2000, \apj, 542, 75
\bibitem[Dall'Aglio, Wisotzki, \& Worseck (2008a)]{dallaglio} Dall'Aglio, A., Wisotzki, L., \& Worseck, G. 2008a, \aap, 480, 359
\bibitem[Dall'Aglio, Wisotzki, \& Worseck (2008b)]{dallaglio2} \underline{\hskip 7em}, 2008b, \aap, 491, 465
\bibitem[Davidsen, Kriss \& Zheng(1996)]{dkz} Davidsen, A. F., Kriss, G. A. \& Zheng, W. 1996, Nature 380, 47
\bibitem[Donahue \& Shull (1987)]{donahue} Donahue, M. \& Shull, J. M.\  1987, \apj, 323, L13
\bibitem[Fan, Carilli \& Keating(2006)]{fan} Fan, X., Carilli, C. L., \& Keating, B., 2006, ARAA, 44, 415
\bibitem[Fardal, Giroux \& Shull (1998)]{fardal} Fardal, M. A., Giroux, M. L., \& Shull, J. M., 1998, \aj, 115, 2206
\bibitem[Fechner et~al.(2004)]{fechner04} Fechner, C., Baade, R., \& Reimers, D. 2004, \aap, 418, 857
\bibitem[Fechner et al. (2006)]{fechner} Fechner, C., Reimers, D., Kriss, G, A. et al. 2006, \aap, 455, 91
\bibitem[Feldman et al. (1992)]{feldman} Feldman, P. D., Davidsen, A. F., Blair, W. P. et al. 1992, Geophy. Rev. Lett., 19, 453
\bibitem[Fitzpatrick (1999)]{ext} Fitzpatrick, E. L. 1999, PASP, 111, 63
\bibitem[Giroux, Fardal \& Shull(1995)]{giroux} Giroux, M. L., Fardal, A. A., \& Shull, J. M. 1995, \apj, 451, 477
\bibitem[Green et al. (2012)]{green} Green, J. C., Froning, C. S., Osterman, S. et al.\ 2012, \apj, 744, 60  
\bibitem[Gunn \& Peterson(1965)]{gp} Gunn, J. \& Peterson, B. 1965, \apj,   142, 1633
\bibitem[Haardt \& Madau(1996)]{haardt} Haardt, F. \& Madau, P. 1996, \apj, 461, 20 
\bibitem[Heap et al.(2000)]{heap} Heap, S. R., Williger, G. M., Smette, A. et al.\ 2000, \apj, 534, 69
\bibitem[Hewett \& Wild(2010)]{hewett} Hewett, P. C. \& Wild, V. 2010, \mnras, 405, 2302 
\bibitem[Hodge (2011)]{hodge} Hodge, P. E. \ 2011, in {\it Astronomical Data Analysis 
Software and Systems XX},  eds. I. N. Evans, A. Accomazzi, D. J. Mink, \&  A. H. Rots, 
({\it A. S. P. Conf. Series 442}, ASP, San Francisco), 391
\bibitem[Hogan, Anderson \& Rugers (1997)]{hogan} Hogan, C. J.,  Anderson, S. F. \&
  Rugers, M. H. 1997, \aj, 113, 1495
\bibitem[Hogg (1999)]{hogg} Hogg, D. W. 1999, astro-ph/9905116
\bibitem[Isobe, Feigelson \& Nelson (1986)]{isobe} Isobe, T., Feigelson, E. D. \& Nelson, P. I.  1986, ApJ, 306, 490
\bibitem[Jakobsen et al. (1994)]{jakobsen} Jakobsen, P., Boksenberg, A., Deharveng, J. M., Greenfield, P., Jedrzejewski, R.,
 \& Paresce, F. 1994, Nature, 370, 35
\bibitem[Janknecht et al.(2006)]{janknecht} Janknecht, E., Reimers, D., Lopez, S. \& Tytler, D. 2006,
\aap, 458, 427
\bibitem[Kim et al.(2002)]{kim2002} Kim, T.-S., Carswell, R. F., Cristiani, S., D'Odorico, S., \& Giallongo, E. 2002, \mnras, 335, 555
\bibitem[Kim et al.(2013)]{kim2013} Kim, T.-S., Partl, A. M., Carswell, R. F., \& M\"uller, V. 2013, \aap, 552, 77
\bibitem[Kirkman et al.(2007)]{kirkman} Kirkman, D., Tytler, D., Lubin, D., \& Charlton, I. 2007, MNRAS, 376, 1227
\bibitem[Komatsu et al.(2011)]{wmap} Komatsu, E., Smith, K. M., Dunkley, J. et al.\ 2011, \apjs, 192, 18
\bibitem[Kriss (1994)]{specfit} Kriss, G. A. 1994, in 
  {\it Astronomical Data Analysis Software and  Systems III}, eds. D. R. Crabtree, 
  R. J.  Hanisch, \& J. Barnes, ({\it A. S. P. Conf. Series 61}, ASP, San Francisco), 437 
\bibitem[Kriss et al.(2001)]{gak} Kriss, G. A., Shull, J. M., Oegerle, W. et al.\ 2001, Science, 293, 1112
\bibitem[Madau \& Meiksin (1994)]{madau94} Madau, P. \& Meiksin, A. 1994, \apj,   433, L53
\bibitem[Madau \& Rees(2000)]{madau00} Madau, P. \& Rees, M. J. 2000, \apj,   542, 69
\bibitem[Martin et al. (2005)]{martin} Martin, D. C., Fanson, J., Schiminovich, D. et al. 2005, \apj, 619, L1
\bibitem[McQuinn et al. (2009)]{mcquinn} McQuinn, M., Lidz, A., Zaldarriaga, M., et al. 2009, \apj, 694, 842
\bibitem[McQuinn \& Worseck (2014)]{mcquinn2014} McQuinn, M. \& Worseck, G. 2014, \mnras, 440, 2406
\bibitem[Meiksin (2005)]{meiksin05} Meiksin, A. 2005, MNRAS, 356, 596  
\bibitem[Meiksin (2006)]{meiksin06} \underline{\hskip 7em} 2006, MNRAS, 365, 807  
\bibitem[Meiksin (2009)]{meiksin09} \underline{\hskip 7em} 2009, Rev. Modern Phys., 841, 1405 (M09) 
\bibitem[Miralda-Escud\'e(1998)]{me} Miralda-Escud\'e, J. 1998, \apj, 501, 15
\bibitem[Mortlock et al. (2011)]{mortlock} Mortlock, D. J., Warren, S. J., Venemans, B. P. et al.  2011, Nature, 474, 616
\bibitem[Morton(1991)]{morton} Morton, D. C, 1991, \apjs, 77, 119
\bibitem[Murdoch et al. (1986)]{murdoch} Murdoch, H. S., Hunstead, R. W., Pettini, M. \& Blades, J. C. 1986, \apj, 309, 19
\bibitem[Osterbrock \& Ferland (2006)]{agn} Osterbrock, D. E. \& Ferland G. J. 2006
Astrophysics of Gaseous Nebulae and Active Galactic Nuclei (University Science Books, Sausalito, CA)
\bibitem[Reimers et al. (1997)]{reimers} Reimers, D., K\"ohler, S., Wisotzki,
  L., Groote, D., Rodriguez-Pascual, P. \& Wamsteker, W. 1997, \aap, 327, 890
\bibitem[Ribaudo, Lehner \& Howk(2011)]{ribaudo} Ribaudo, J., Lehner, N. \& Howk, J. C. 2011, \apj,  736, 42 \bibitem[Richards et al.(2006)]{richards} Richards, G. T., Strauss, M. A., Fan, X. et al. 2006, \aj, 131, 2766 
\bibitem[Schneider et al.(2010)]{dr7} Schneider, D. P., Richards, G. T., Hall, P. B. et al.\ 2010, \aj, 139, 2360
\bibitem[Scholz \& Walters (1991)]{scholz} Scholz, T. T. \& Walters, H. R. J.\ 1991, \apj, 380, 302  
\bibitem[Shapiro (1986)]{shapiro} Shapiro, P. R. \ 1986, \pasp, 98, 1014
\bibitem[Shull et al.(2010)]{shull} Shull, J. M., France, K., Danforth, C. W. et al.\ 2010, \apj, 722, 1312
\bibitem[Shull et al.(2004)]{shull04} Shull, J. M., Tumlinson, J. Giroux, M. L. et al.\ 2004, \apj, 600, 570 
\bibitem[Smette et al.(2002)]{smette} Smette, A., Heap, S. R., Willinger, G. M., Tripp, T. M., Jenkins, E. B., \& Songalia, A. \ 2002, \apj, 564, 542
\bibitem[Steinhardt \& Elvis(2011)]{se} Steinhardt, C. L. \& Elvis, M. 2011, \mnras, 410, 201
\bibitem[Stengler-Larrea et al. (1995)]{esl}Stengler-Larrea, E. A. Boksenberg, A., Steidel, C. C. et al. 1995, ApJ, 444, 64
\bibitem[Storrie-Lombardi  et al.(1994)]{lsl}Storrie-Lombardi, L. J., McMahon, R. G., Irwin, M. J., \&  Hazard, C. 1994, ApJ, 427, L13 
\bibitem[Str\"omgren(1939)]{stromgren} Str\"omgren, B. 1939, \apj, 89, 526
\bibitem[Syphers (2010)]{syphers} Syphers, D. 2010, PhD dissertation, University of Washington
\bibitem[Syphers et al.(2009{\natexlab{a}})]{syphers0} Syphers, D., Anderson, S. F., Zheng, W. et al.\ 2009\natexlab{a}, \apjs, 185, 20
\bibitem[{Syphers} {et~al.}(2009{\natexlab{b}})]{syphers1} \underline{\hskip 7em} 2009{\natexlab{b}}, \apj, 690, 1181
\bibitem[Syphers et al.(2011{\natexlab{a}})]{syphers2} \underline{\hskip 7em} 2011\natexlab{a}, \apj, 726, 111
\bibitem[Syphers et al.(2011{\natexlab{b}})]{syphers3} \underline{\hskip 7em} 2011\natexlab{b}, \apj, 742, 99
\bibitem[Syphers et al.(2012)]{syphers4} \underline{\hskip 7em} 2012,  \aj, 143, 100
\bibitem[Syphers \& Shull(2013)]{syphers5} Syphers, D. \& Shull, J. M. 2013,  \apj, 765, 119
\bibitem[Syphers \& Shull (2014)]{syphers6} \underline{\hskip 7em}  2014, \apj, 784, 42
\bibitem[Telfer et al.(2002)]{telfer} Telfer, R.,  Zheng, W., Kriss, G. A., \& Davidsen, A. F.\ 2002, \apj, 565, 733 
\bibitem[Tittley \& Meiksin (2007)]{tittley} Tittley, E. R., \& Meiksin, A. 2007, \mnras, 380, 1369
\bibitem[Trainor \& Steidel (2012)]{trainor} Trainor, R. F. \& Steidel, C. C.  2012, \apj, 752, 39 
\bibitem[Tytler(1987a)]{tytler} Tytler, D. 1987a, \apj, 321, 49
\bibitem[Tytler(1987b)]{tytler2} \underline{\hskip 7em} 1987b, \apj, 321, 69 \bibitem[Vanden Berk et al. (2001)]{vb} Vanden Berk, D. E., Richards, G. T., Bauer, A. et al. 2001, \aj, 122, 549
\bibitem[V\'eron-Cetty \& V\'eron (2010)]{veron} V\'eron-Cetty, M.-P. \& V\'eron, P. 2010, \aap, 518A, 10
\bibitem[Venemans et al. (2015)]{ven2} Venemans, B. P., Ba\~nados, E., Decarli, R. et al. 2015, arXiv 15020.1927
\bibitem[Venemans et al. (2013)]{ven1} Venemans, B. P., Findlay, J. R., Sutherland, W. J.  et al. 2013, \apj, 779, 24
\bibitem[Vestergaard \& Peterson (2006)]{vp} Vestergaard, M., \& Peterson, B. 2006, \apj, 641, 689
\bibitem[White et al. (2003)]{white} White, R. L., Becker, R., Fan, X., \& Strauss, M. A. 2003, \aj, 126, 1
\bibitem[Wilson et al. (2004)]{wilson} Wilson, J. C., Henderson, C. P., Herter, T. L. et al. 
2004, in Ground-based Instrumentation for Astronomy, Proc. SPIE 5492, eds. A. F. M. Moorwood \& M. Iye, 1295 
\bibitem[Worseck et al. (2014)]{worseck14} Worseck, G., Prochaska, J. X., Hennawi, J. F. \& McQuinn, M. 2014, arXiv 1405.7405
\bibitem[Worseck et al. (2011)]{worseck} Worseck, G., Prochaska, J. X., McQuinn, M. et al. 2011, \apj, 733, 24
\bibitem[Wu et al. (2015)]{wu} Wu, X., Wang, F., Fan, X. et al. 2015, Nature, 518, 512 
\bibitem[York et al. (2000)]{york} York, D. G., Adelman, J., Anderson, J. E. Jr. et al. 2000, \aj, 120, 1579
\bibitem[Zheng \& Davidsen (1995)]{zheng95} Zheng, W. \& Davidsen, A. F. 1995, \apj, 440, L53
\bibitem[Zheng et al.(2004)]{zheng04} Zheng, W., Kriss, G. A., Deharveng, J.-M.\ et al.\ 2004, \apj, 605, 631
\bibitem[Zheng et al. (1997)]{zheng97} Zheng, W., Kriss, G. A., Telfer, R. C., Grimes, J. P. \& Davidsen, A. D. 1997, \apj, 475, 469
\bibitem[Zheng et al.(2008)]{zheng08} Zheng, W., Meiksin, A., Pifko, K. \ et al.\ 2008, \apj, 686, 195
\end{thebibliography}
\end{document}